\documentclass{JHEP3}

\usepackage{amsmath,amssymb, amsfonts}
\usepackage{graphicx}
\usepackage{epsfig}

\def\braket#1{\mathinner{\langle{#1}\rangle}}
\newcommand{\sbraket}[1]{\lbrack #1\rbrack}

\newcommand{\dalpha}{\dot{\alpha}}

\newcommand{\tr}{\textrm{Tr}}

\newcommand{\ii}{\mathrm{i}}
\newcommand{\al}{{\alpha'}}
\newcommand{\id}{\textrm{I}}
\newcommand{\half}{\frac{1}{2}}

\title{No triangles on the moduli space of maximally supersymmetric gauge theory}
\author{Rutger H. Boels \\ Niels Bohr International Academy and DISCOVERY center, Niels Bohr Institute\\ Blegdamsvej 17, DK-2100 Copenhagen, Denmark}


\abstract{Maximally supersymmetric gauge theory in four dimensions has a remarkably simple S-matrix at the origin of its moduli space at both tree and loop level. This leads to the question what, if any, of this structure survives at the complement of this one point. Here this question is studied in detail at one loop for the branch of the moduli space parameterized by a vacuum expectation value for one complex scalar. Motivated by the parallel D-brane picture of spontaneous symmetry breaking a simple relation is demonstrated between the Lagrangian of broken super Yang-Mills theory and that of its higher dimensional unbroken cousin. Using this relation it is proven both through an on- as well as an off-shell method there are no so-called triangle coefficients in the natural basis of one-loop functions at any finite point of the moduli space for the theory under study. The off-shell method yields in addition absence of rational terms in a class of theories on the Coulomb branch which includes the special case of maximal supersymmetry. The results in this article provide direct field theory evidence for a recently proposed exact dual conformal symmetry motivated by the AdS/CFT correspondence.}

\begin{document}

\section{Introduction and motivation}
Recent years have seen the accumulation of evidence that S-matrices in four dimensions display hidden symmetries which are not obvious from their original off-shell formulation. This is interesting from the point of view of applications to real-world physics as amplitudes form a direct bridge between (quantum field) theory and experiment. More theoretically, there is the hope that unraveling perturbative symmetries will lead to a non-perturbative understanding of gauge theory. Especially the latter motivation is relevant for studying maximally supersymmetric Yang-Mills theory as here a beginning of such an understanding is given by the AdS/CFT correspondence \cite{Maldacena:1997re}.

Using this correspondence Alday and Maldacena \cite{Alday:2007hr} discussed scattering amplitudes from a strong coupling perspective for unbroken $\mathcal{N}=4$ SYM in four dimensions. Their work suggested the relevance of certain light-like Wilson loops to perturbative calculations of planar MHV amplitudes. Remarkably this link was also verified on the weak coupling side \cite{Drummond:2007aua, Brandhuber:2007yx}. The Wilson loop picture suggests a new `dual' superconformal symmetry of $\mathcal{N}=4$ at the origin of the moduli space \cite{Drummond:2008vq} (see \cite{Henn:2009bd} and references therein for a more complete discussion). This new symmetry appears in addition to the more familiar conformal symmetry and is conjectured to be broken at one loop only in dimensional regularization with the anomaly explicitly known (see \cite{Brandhuber:2009kh} and references therein). Recently it was suggested \cite{Alday:2009zm} that with a different regularization the dual conformal symmetry might be made exact. More evidence for this was presented in \cite{Henn:2010bk}. The main idea of this regularization is to move away from the origin of the moduli space to a generic point where one of the scalars acquires a vev, breaking the gauge symmetry through the Higgs mechanism. This is sometimes referred to as the Coulomb branch of the moduli space.

A problem with amplitudes on the Coulomb branch is that most of the recently developed technology for analytic scattering amplitude calculations has focused on massless particles first and foremost. This paper is part of an effort to address this. This is especially important since massive particles are after all ubiquitous in Nature. Of particular concern here is the one-loop field theory verification of the prediction motivated from a strong coupling analysis through AdS/CFT in \cite{Alday:2009zm} that there are no so-called triangle coefficients in a natural basis of functions for one loop amplitudes on the Coulomb branch as these would be incompatible with the dual conformal symmetry in the string theory setup. Hence showing the absence of triangles in the field theory at weak coupling directly is evidence that such an exact symmetry exists perturbatively, at least to one loop order.

Apart from AdS/CFT motivations, the question how amplitudes depend on moduli space parameters is certainly interesting in its own right. With this motivation the general absence of triangles was speculated about first in \cite{Schabinger:2008ah}. In a sense this absence is an extension of the physical intuition that modifications to the IR of the theory should not influence the UV behavior. Usually this intuition is made precise and studied for the UV divergent contributions with a view towards renormalisation; here it will be shown to hold also for the `sub-divergent' part of the theory under study. Note that the intuition does immediately rule out bubble and tadpole graphs anywhere on the Coulomb branch of $\mathcal{N}=4$ as these would be UV divergent.

This article is structured as follows: first some of the pieces of the puzzle will be introduced in section \ref{sec:puzzlepieces}. Some results from the literature are reproduced here for convenience of reader and author; new here is for instance an analysis of BCFW shifts of massive vector boson legs and the massive spinor helicity interpretation of three particle cuts. In section \ref{sec:stringpov} the string theory picture of gauge boson amplitudes will be extended to open strings stretched between parallel separated branes, building on the vertex operators found by \cite{Pesando:1999hm}. This picture yields in the field theory limit a direct map between amplitudes in higher dimensional gauge theory and spontaneously broken gauge theory in lower dimensions. The same result will also be derived purely within field theory on the level of the Lagrangian in subsection \ref{sec:fieldtheorypov}. With this final piece of information in hand two versions of a no-triangle argument will be set up in section \ref{sec:notriangles}. On-shell the argument will be along the same lines as was done for $\mathcal{N}=4$ SYM at the origin of the moduli space in \cite{ArkaniHamed:2008gz}. Off-shell the argument very closely resembles \cite{Bern:1994zx}. Moreover, the latter yields absence of rational terms. The presentation is capped of by a discussion and conclusion section. In appendix \ref{app:bcfwgenspinax} BCFW shifts for massive vector bosons with a common axis are analyzed. Further, in appendix \ref{app:susyrepsBPS} the on-shell superspace for BPS representations of $\mathcal{N}=4$ SYM is constructed directly in four dimensions by a specialization of the higher dimensional approach of \cite{Boels:2009bv}. Finally, in appendix \ref{app:savingphi4} it is shown how  $\phi^4$ theory can be made to evade some slanderous comments.

\section{Kinematics of massive legs up to one loop}\label{sec:puzzlepieces}

\subsection{Spinor helicity and on-shell recursion for massive states}
Massive spinor helicity in four dimensions is discussed in \cite{Dittmaier:1998nn}. The basic formula needed to start the analysis is
\begin{equation}\label{eq:decomponemom}
k_{\mu} = k^{\flat}_{\mu} + \frac{k^2}{2 q \cdot k} q_{\mu}
\end{equation}
for any massive momentum vector $k$ in terms of two light-like momenta $q$ and $k^\flat$. From consistency of this equation the condition
\begin{equation}
q \cdot k = q \cdot k^{\flat}
\end{equation}
follows which makes the equation above invertible. The formula can be made transparent by drawing a lightcone diagram in an arbitrary frame as in figure \ref{fig:onemomdecomp}.

\DOUBLEFIGURE[ht]{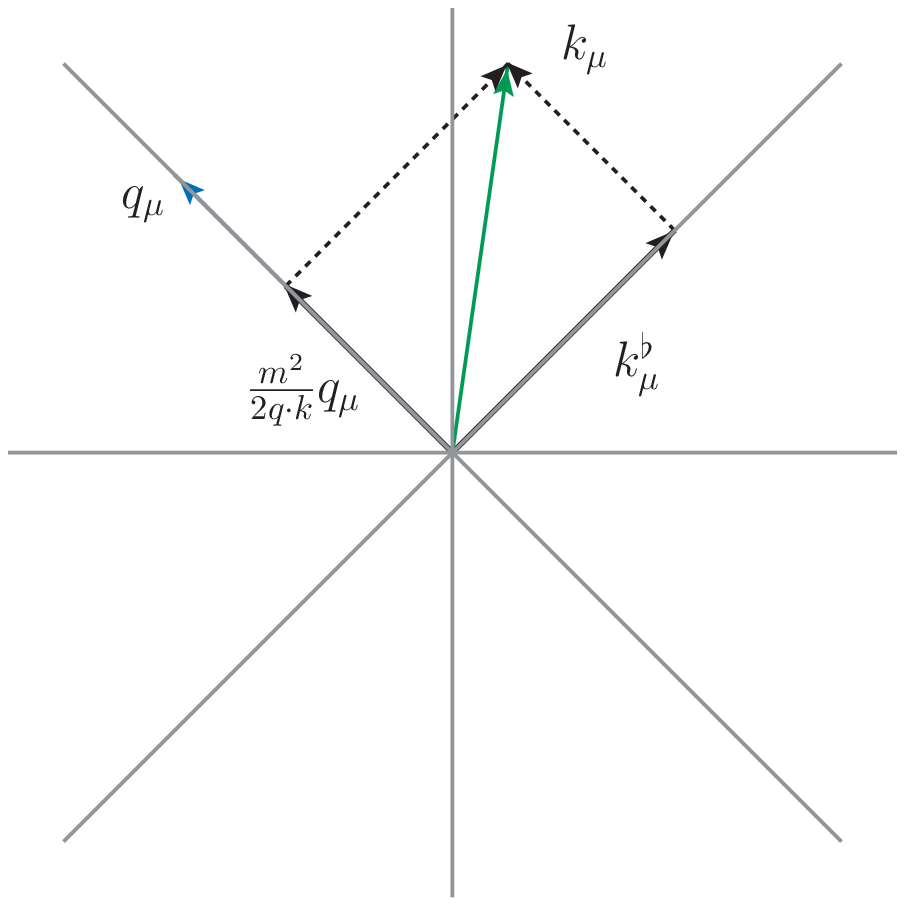,scale=0.6}{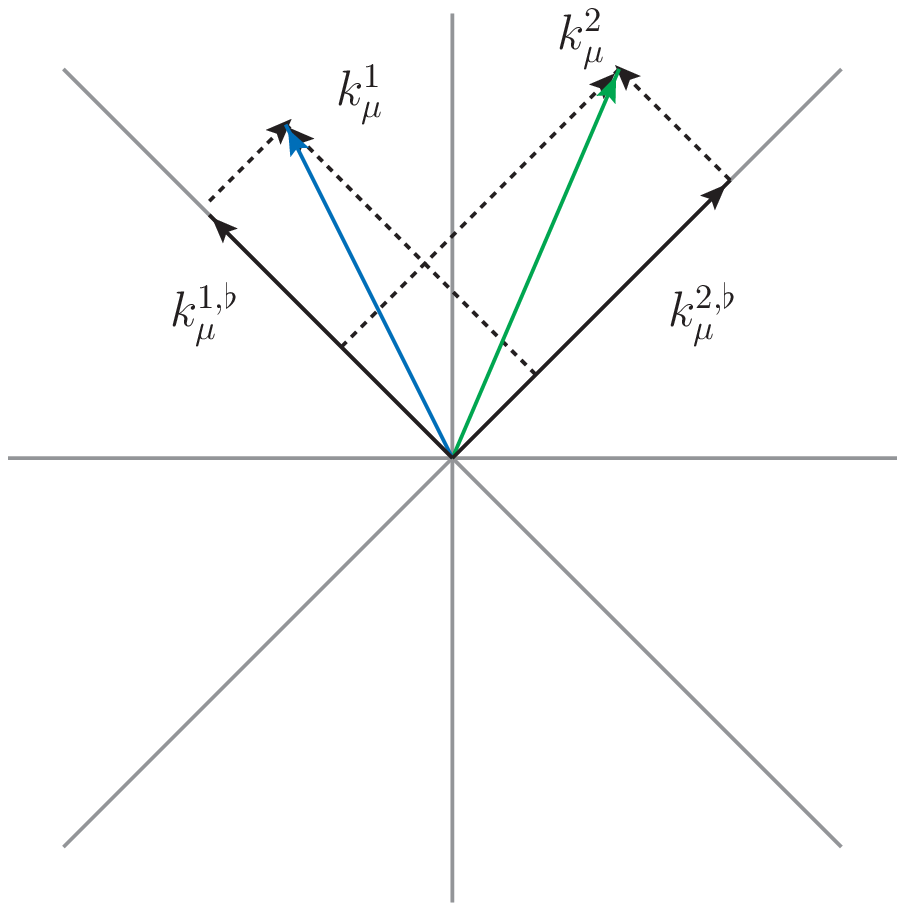,scale=0.6, }{Decomposing a time-like vector into two light-like ones \label{fig:onemomdecomp}}{Decomposing two time-like vectors into light-like ones \label{fig:twomomdecomp}}

To see that a choice of $q$ is necessary one can transform to the rest-frame for one particle. In this frame a decomposition into light-like vectors is ambiguous: $q$ fixes a particular choice of space-like axis. As will be done throughout this article, it is natural to take this axis to be the spin polarization axis, isolated covariantly as
\begin{equation}\label{eq:defspinaxis}
s_{\mu} = k_{\mu} - \frac{k^2}{q \cdot k} q_{\mu}
\end{equation}
The natural quantum numbers are the eigenvalues under rotations around this axis. The corresponding operator is
\begin{equation}\label{eq:paulilubanski}
R_z = \frac{q_{\mu} W^{\mu}}{2 q \cdot k} \equiv  \frac{q_{\mu} k_{\nu} \Sigma_{\rho \sigma} \epsilon^{\mu\nu\rho\sigma} }{2 q \cdot k}
\end{equation}
with $W_{\mu}$ the Pauli-Lubanski vector and $\Sigma$ the rotation generator of the Lorentz group. This can be checked straightforwardly in the rest-frame. Note that in the massless limit the spin quantum number smoothly reduces to helicity.

For two massive momenta a similar drawing to the case of one massive momentum can be made, see figure \ref{fig:twomomdecomp}. The analog of equation \eqref{eq:decomponemom} for two massive vectors $k_1$ and $k_2$ reads,
\begin{eqnarray}\label{eq:massivemomentumexp}
k_1 = k^{\flat}_1 + \frac{m_1^2}{\gamma_{12}} k^{\flat}_2 & \quad & k_2 = k^{\flat}_2 + \frac{m_2^2}{\gamma_{12}} k^{\flat}_1
\end{eqnarray}
in terms of two massless momenta $k^{\flat}_1$ and $k^{\flat}_2$. From consistency of this decomposition the condition
\begin{align}
\gamma_{12} & = 2 (k^{\flat}_1 \cdot k^{\flat}_2) = \left((k_1 \cdot k_2) \pm \sqrt{(k_1 \cdot k_2)^2 - m_1^2 m_2^2}\right)
\end{align}
follows. For a smooth massless limit for real momenta the $+$ sign should be chosen. The figure above also illustrates that the ambiguity present for one massive momentum is absent in the case of two momenta.

\subsubsection*{Basis of vectors}
For two given massive momenta there is a natural `lightcone' basis for all vectors spanned by
\begin{eqnarray}\label{eq:basisvecsspinhel}
a^{1,\alpha \dalpha} = k_1^{\flat, \alpha} k_1^{\flat, \dalpha} & \quad & a^{3,\alpha \dalpha} = k_1^{\flat, \alpha} k_2^{\flat, \dalpha} \nonumber \\
a^{2,\alpha \dalpha} = k_2^{\flat, \alpha} k_2^{\flat, \dalpha} & \quad & a^{4,\alpha \dalpha} = k_2^{\flat, \alpha} k_1^{\flat, \dalpha}
\end{eqnarray}
in terms of the Weyl spinors associated to the massless momenta in \eqref{eq:massivemomentumexp}. Note that in particular,
\begin{equation}
a^{3,\mu} a^4_{\mu} = - a^{1,\mu} a^2_{\mu}
\end{equation}
In the following the notation
\begin{equation}
k_1^{\flat, \dalpha} \equiv 1^{\dalpha} \qquad  k_1^{\flat, \alpha}  \equiv 1^{\alpha} \qquad k_2^{\flat, \dalpha}  \equiv 2^{\dalpha} \qquad k_2^{\flat, \alpha}  \equiv 2^{\alpha}
\end{equation}
will be employed frequently.

One way to to interpret this basis is to observe that there is a frame in which these basis vectors have the form
\begin{equation}\label{eq:choiceofframe}
\begin{array}{cccc} a^1_{\mu} \sim  \left(\begin{array}{c} 1 \\ 0 \\ 0 \\ 1 \end{array}\right) \quad &
a^2_{\mu} \sim  \left(\begin{array}{c} 1 \\ 0 \\ 0 \\ -1 \end{array} \right) \quad  &
a^3_{\mu} \sim \left(\begin{array}{c} 0 \\ 1 \\ i \\ 0 \end{array}\right) \quad  &
a^4_{\mu} \sim  \left(\begin{array}{c} 0 \\ 1 \\ -i \\ 0 \end{array} \right)
\end{array}
\end{equation}
up to a trivial normalization. The spinors of \eqref{eq:basisvecsspinhel} are the covariant version of this observation. Note that in principle any of the four massless vectors in the basis can be chosen to define the spin polarization axis. An arbitrary momentum vector $l$ can be expanded into the basis as
\begin{align}\label{eq:momentumexpansion}
l_{\mu} & = c_1 a^{1}_{\mu} + c_2 a^{2}_{\mu} + c_3 a^{3}_{\mu} + c_4 a^{4}_{\mu} \\
& = c_1 a^{1}_{\mu} + c_2 a^{2}_{\mu} + c_4 a^{4}_{\mu} + \left(c_3 - \frac{l^2}{2 a^3 \cdot l} \right) a^{3}_{\mu} + \frac{l^2}{2 a^3 \cdot l} a^3_{\mu} \\
& \equiv c_1 a^{1}_{\mu} + c_2 a^{2}_{\mu} + c_4 a^{4}_{\mu} + \tilde{c}_3  a^{3}_{\mu} + \frac{l^2}{2 a^3 \cdot l} a^3_{\mu} \label{eq:defoflflat} \\
& \equiv l^{\flat}_{\mu} + \frac{l^2}{2 a^3 \cdot l} a^3_{\mu}
\end{align}
One way to proceed would be to translate to spinor helicity expressions in the frame given above, e.g.
\begin{equation}
l^{\flat}_{\mu} \sigma^{\mu} = \left(\begin{array}{cc} c_1 & c_4 \\ \left(c_3 - \frac{l^2}{2 a^3 \cdot l} \right) & c_2 \end{array}\right) \qquad \qquad \textrm{(in special frame)}
\end{equation}
The determinant of this expression (equivalent to $l^{\flat}\cdot l^{\flat}$) vanishes by construction and up to the usual scaling ambiguity the spinors associated to $l^{\flat}$ can be found by a simple direct calculation. See Appendix \ref{app:bcfwgenspinax} for more on this. It is more elegant however to identify the coefficients in equation \eqref{eq:defoflflat} with the spinor inner products with the basis vectors. With the massless momenta $l^{\flat}$ written in terms of two Weyl spinors as
\begin{equation}\label{eq:scalambig}
l^{\flat}_{\alpha \dalpha} = \lambda_\alpha \lambda_{\dalpha}
\end{equation}
one can obtain
\begin{equation}\label{eq:getspinorinnerprods}
\begin{array}{cc}
c_1  =  \frac{\braket{2 \lambda}\sbraket{2 \lambda}}{\gamma_{12}}  &  c_2  =  \frac{\braket{1 \lambda}\sbraket{1 \lambda}}{\gamma_{12}} \\
c_4  =  \frac{\braket{2 \lambda}\sbraket{1 \lambda}}{\gamma_{12}}  &  \tilde{c}_3  =  \frac{\braket{1 \lambda}\sbraket{2 \lambda}}{\gamma_{12}}
\end{array}
\end{equation}
with
\begin{equation}
\gamma_{12} = \braket{12}\sbraket{12}
\end{equation}
Therefore, given the basis coefficients $c_i$ one can obtain through \eqref{eq:getspinorinnerprods} the spinor inner products of $l^{\flat}$ with a complete basis of dotted and undotted Weyl spinors. This allows one to obtain the actual spinors from
\begin{align}
\lambda_{\alpha} & = \frac{\braket{\lambda 2} 1_{\alpha} - \braket{\lambda 1} 2_{\alpha} }{\braket{1 2}} \\
\lambda_{\dalpha} & = \frac{\sbraket{\lambda 1} 2_{\dalpha} - \sbraket{\lambda 2} 1_{\dalpha}}{\sbraket{1 2}}
\end{align}
which simply express the fact that Weyl spinors in four dimensions are two dimensional. The degree of freedom counting works out if one bears in mind that equations like \eqref{eq:scalambig} only determine spinors up to a scaling.

\subsection*{Polarization vectors and BCFW shifts}
In unitary gauge the polarization states of the massive vector boson with the above choice of spin quantization axis of equation \eqref{eq:defspinaxis} are given as
\begin{equation}\label{eq:polvecsmas}
e^+_{\alpha \dalpha} = \frac{1}{\sqrt{2}} \frac{k^{\flat}_{\alpha} q_{\dalpha}}{\sbraket{q k^{\flat}}} \quad e^-_{\alpha \dalpha} = \frac{1}{\sqrt{2}} \frac{q_{\alpha} k^{\flat}_{\dalpha}}{\braket{q k^{\flat}}} \quad e^{0}_{\alpha \dalpha} = \frac{s_{\alpha \dalpha}}{m}
\end{equation}
in this equation $s_{\mu}$ is the spin quantization axis given above in equation \eqref{eq:defspinaxis}.

For two massive particles there is a special choice of spin axes which deserves some special mention. This is the choice for which the spin quantization axis of the first particle is determined by the massless projection of the momentum of the second and vice-versa. In formulas the massless vector $q^i$ for leg $i$ is given as
\begin{equation}
q^1 = k^{2,\flat}_{\mu} \qquad q^2 = k^{1,\flat}_{\mu}
\end{equation}
The polarization vectors follow from \eqref{eq:polvecsmas} for both legs and are related as
\begin{equation}\label{eq:relationbetweenvecs}
e^{\pm}_1 = - e^{\mp}_2
\end{equation}
From the definition \eqref{eq:paulilubanski} it is easy to see that in general the quantum numbers of the states are inverted between the two legs ($h$ $\rightarrow$ $-h$). Note the close relation to the center of mass frame by the relation to equation \eqref{eq:basisvecsspinhel}.

The machinery of this subsection has a neat application to the derivation of on-shell recursion relations through BCF(W) shifts \cite{Britto:2004ap} (\cite{Britto:2005fq}) of massive vector (and tensor) particles which may be of interest independent of the present article. Shifts of amplitudes which involve massive particles were first discussed briefly in \cite{Badger:2005zh}. In brief, BCFW turn a given amplitude $A$ into function of a complex variable $z$ by deforming two of the momenta
\begin{equation}
k_i \rightarrow k_i + n z \qquad k_i \rightarrow k_i - n z
\end{equation}
for a vector $n$ for which
\begin{equation}\label{eq:BCFWconstraintsn}
k_i \cdot n = k_j \cdot n = n \cdot n = 0
\end{equation}
This gives $A \rightarrow A(z)$. In the BCFW argument, certain on-shell recursion relations are proven if the following is true at tree level,
\begin{equation}
Res_{z = \infty} \frac{A(z)}{z} =0
\end{equation}
A sufficient condition for this is $A(z) \rightarrow 0$ which has been proven in a variety of field theories, see e.g. \cite{ArkaniHamed:2008yf}, \cite{Cheung:2008dn} and references therein.

For two four-dimensional massless momenta the vector $n$ is easily constructed and used for applications. For massive momenta, a covariant solution can be given in terms of the polarization vectors defined above: just take $n$ to be proportional to either $e^{+}$ or $e^{-}$,
\begin{equation}
n_{\alpha \dalpha} = k^{1,\flat}_{\alpha} k^{2,\flat}_{\dalpha} \quad \textrm{or} \quad n_{\alpha \dalpha} = k^{2,\flat}_{\alpha} k^{1,\flat}_{\dalpha}
\end{equation}
This solution can easily seen to be correct in the special frame given in equation \eqref{eq:choiceofframe}. These are the two independent solutions to equations \eqref{eq:BCFWconstraintsn}. This solution is also used in \cite{Schwinn:2007ee} where shifts of massive quark lines are discussed. Here we will focus on shifting massive vector boson legs. The polarization vectors for the deformed momenta can be obtained straightforwardly for, say, the first of these solutions. The only vector which is not invariant changes as
\begin{equation}\label{eq:specpolaxshifts}
\begin{array}{ccc}
(-  e_{j}^- =) e_i^+ &\rightarrow  & e_i^+ + z \frac{(k_i^{\flat})^{\flat}_{\alpha}(k_j^{\flat})_{\dalpha}}{\braket{k_i^{\flat} k_j^{\flat}}}
\end{array}
\end{equation}
The other polarization vectors $e_i^- = -  e_{j}^+ $ are proportional to the shift vector $n$ itself. This is very similar to non-covariant higher dimensional analysis of the massless case considered in \cite{ArkaniHamed:2008yf}.

With the above analysis of the polarization vectors simple general bounds can be given for shifting two massive vector bosons in a minimally coupled spontaneously broken gauge theory, summarized in table \ref{tab:BCFWkinmasvec}. This table is produced simply by tracking $z$ dependence in Feynman diagrams in Feynman-'t Hooft gauge, just as was done for massless gluons in the original BCFW paper \cite{Britto:2005fq}. Important here is that the broken gauge theory obeys a Ward identity which implies that at the tree level we are interested the following correlator vanishes,
\begin{equation}
\braket{\left(k_{\mu} A^{\mu} + m \Phi \right) X } = 0
\end{equation}
Here $X$ stands for all other fields in the amplitude and $\Phi$ is the scalar field which acquires a vev. For a shifted momentum this reads
\begin{equation}
\braket{\left((k_{\mu} + z n) A^{\mu} + m \Phi \right)) X } = 0
\end{equation}
Since $e_i^- \sim e_j^+ \sim n$ this can be used to estimate that these legs contribute $\sim \frac{1}{z}$. Working through the possibilities yields table \ref{tab:BCFWkinmasvec}.

\begin{table}
\begin{center}
\begin{tabular}{c|c c c}
$e_i \;\backslash \;e_j  $ & $-$   & $+$   &  $0$     \\
\hline
$-$                        & $ -1$ & $ 1$  &  $0$       \\
$+$                        & $ -3$ & $-1$  & $-2$       \\
$0$                        & $ -1$ & $ 0$  & $-1$
\end{tabular}
\caption{upper bounds on the leading power in $z^{-\kappa}$ in the large $z$ limit for two shifted polarized massive vector bosons in gauge theories with at most one derivative in the vertices.}
  \label{tab:BCFWkinmasvec}
\end{center}
\end{table}

This table of bounds derived by Feynman graphs is the same that obtained in unbroken Yang-Mills theory by this particular method. In unbroken Yang-Mills theory it is known by other methods that for instance the $(++)$ and $(--)$ are in fact better behaved (as $\sim \frac{1}{z}$), while the other two cases are saturated. It is easy to speculate that a similar conclusion holds for spontaneously broken theories in general. Below this will be proven to be true in an important class of particular examples.

The above is for a particular choice of polarization axes which depends on the momenta of the particles involved. An immediate question is if a similar conclusion would hold by shifting two massive vector boson particles with a shared polarization axis defined by some light-like vector $q$. In appendix \ref{app:bcfwgenspinax} this is shown to be true by specializing to a particular Lorentz frame. It should not be too surprising that BCFW holds for spontaneously broken gauge theory as the infinite complex momentum behavior should be linked to the UV properties of the theory under study. These are (famously) unchanged under symmetry breaking which is after all an IR modification. As an example of the opposite case, a lesson that modifying the UV changes behavior under BCFW shifts is contained in appendix \ref{app:savingphi4}.

\subsubsection*{Six and four dimensions}
On-shell massive vector boson states in four dimensions can be interpreted as higher dimensional massless states and this point of view will be useful below. A complete treatment of spinor helicity in higher dimensions (including supersymmetry) is given in \cite{Boels:2009bv}, here only the six dimensional case will be needed. For the massless case in six dimensions the techniques in \cite{Boels:2009bv} are a specific realization of the general forms proposed in \cite{Cheung:2009dc}. The first step involves a choice of orthonormal frame spanned by $q, \bar{q}, n_i$ for $i=1,\ldots,4$ for which the only nontrivial inner products are
\begin{equation}
q\cdot \hat{q} = 1 =  n_j \cdot n_j \qquad \textrm{no sum}
\end{equation}
Without loss of generality $n_3$ and $n_4$ can be taken to span two auxiliary dimensions to the four of interest. This implies
\begin{equation}
R^1_q = q^{\mu} n_1^{\nu} n_2^\rho W_{\mu\nu\rho} \equiv R_z
\end{equation}
where $W_{\mu\nu\rho}$ is the Pauli-Lubanski \emph{tensor}
\begin{equation}
W_{\mu\nu\rho} = k_{[\mu} \Sigma_{\nu \rho]}
\end{equation}
where the brackets denote anti-symmetrization and $R_z$ is the spin generator for the four dimensions spanned by $q, \hat{q}, n_1, n_2$ as in equation \eqref{eq:paulilubanski}. In the `four dimensional limit' given by
\begin{equation}
k \cdot n_3 = k \cdot n_4  \rightarrow 0 \qquad \textrm{four dimensional limit}
\end{equation}
this is the helicity generator. In the same limit the other Cartan sub-algebra generator generates rotations in the plane spanned by $n_3$ and $n_4$. These formulas are all on the level of the algebra, so hold for any representation.

Representations are labeled by the eigenvalues of the Cartan subalgebra of the little group. In six dimensions the little group has rank 2, so there are two labels. The first will be taken to be the eigenvalue under $R_z$, the second the rotation around our chosen $n_3$ and $n_4$. Explicit polarization vectors and spinors can be found in \cite{Boels:2009bv} in a suitable (lightcone) gauge. These have a smooth massless limit, in contrast to those in equation \eqref{eq:polvecsmas} whose gauge is ill-defined in this limit.

\subsection{Using the on-shell superspace for $\mathcal{N}=4$ BPS multiplets}
In this article $\mathcal{N}=4$ SYM theory will be studied with a non-trivial vev for one of the adjoint scalars which will give a mass to some of the vector bosons through the Higgs mechanism. From the point of view of on-shell supersymmetry representation theory, the only way in which an on-shell multiplet can acquire a mass while maintaining the number of degrees of freedom of a `short' representation is by satisfying the BPS condition (see e.g. \cite{Fayet:1978ig, Sohnius:1985qm}). In this subsection an on-shell superspace will be presented for these multiplets. This follows quite straightforwardly from the $D=6$ case of the higher dimensional on-shell superspaces for massless multiplets constructed in \cite{Boels:2009bv}. The `extra' dimensional momenta play the role of central charges which turn into the mass as,
\begin{equation}
m \equiv Z_1 + \ii Z_2 = (k \cdot n_3) + \ii (k \cdot n_4)
\end{equation}
and its conjugate. A four dimensional version of the construction of the on-shell superspace is presented in Appendix \ref{app:susyrepsBPS}. The upshot of the analysis is that there is an on-shell superspace for BPS representations which smoothly generalizes the massless four dimensional case. This arises as a coherent state representation of the fermionic harmonic oscillator and depends on a choice of top state. For the choice of the spin-up polarization of the vector boson state as a top state we have
\begin{equation}
|\,\eta_I, \iota_I \rangle = e^{\sum_I \eta_I Q^I_{-} + \iota_I \overline{Q}^I_-}|\!\uparrow\,\rangle
\end{equation}
and for the spin-down version,
\begin{equation}
|\,\overline{\eta}_I, \overline{\iota}_I \rangle = e^{\sum_I \overline{\eta}_I Q^I_{+} + \overline{\iota}_I \overline{Q}^I_+}|\!\downarrow\, \rangle
\end{equation}
Here $I$ runs from one to two and subscripts on the susy generators indicate raising and lowering of the spin quantum number. The variables $\eta_I, \overline{\eta}_I,\iota_I$ and  $\overline{\iota}_I$ are fermionic. Expanding in the fermionic variables in both cases produces a complete $\mathcal{N}=4$ BPS multiplet. As derived in equation \eqref{eq:trafosusycohstateDirac} and \eqref{eq:trafosusycohstateIIDirac} finite supersymmetry transformations given by a Dirac spinor $\xi$ or its conjugate read
\begin{align}\label{eq:trafosusycohstateWeyl}
e^{\overline{\xi}_I Q_I} |\,\eta_I, \iota_I \rangle & = e^{\iota_I \left(m \frac{\sbraket{\overline{\xi}_I q}}{\sbraket{q k^{\flat}}} + \braket{\overline{\xi}_I k^{\flat}}\right)} \, \,|\,\eta_I + \left(\sbraket{\overline{\xi}_I k^{\flat}} + \overline{m} \frac{\braket{\overline{\xi}_I q}}{\braket{q k^{\flat}}} \right), \iota_I \rangle \\
e^{\overline{Q}_I \xi_I} |\,\eta_I, \iota_I \rangle & = e^{\eta_I \left(m \frac{\sbraket{q \xi_I}}{\sbraket{q k^{\flat}}} + \braket{k^{\flat} \xi_I} \right)}\,\,|\,\eta_I, \iota_I + \left(\sbraket{k^{\flat} \xi_I} + \overline{m} \frac{\braket{q \xi_I}}{\braket{q k^{\flat}}}  \right) \rangle
\end{align}
and
\begin{align}\label{eq:trafosusycohstateIIWeyl}
e^{\overline{\xi}_I Q^I} |\,\overline{\eta}_I, \overline{\iota}_I \rangle & = e^{\overline{\iota}_I \left(\sbraket{\overline{\xi}_I k^{\flat}} + \overline{m} \frac{\braket{\overline{\xi}_I q}}{\braket{q k^{\flat}}}\right) }\,\,|\,\overline{\eta}_I + \left(m \frac{\sbraket{\overline{\xi}_I q}}{\sbraket{q k^{\flat}}} + \braket{\overline{\xi}_I k^{\flat}}\right), \overline{\iota}_I \rangle \\
e^{\overline{Q}_I \xi^I} \,\,|\,\overline{\eta}_I, \overline{\iota}_I \rangle & = e^{\overline{\eta}_I \left(\sbraket{k^{\flat} \xi_I} + \overline{m} \frac{\braket{q \xi_I}}{\braket{q k^{\flat}}} \right) }|\,\overline{\eta}_I, \overline{\iota}_I + \left(m \frac{\sbraket{q \xi_I}}{\sbraket{q k^{\flat}}} + \braket{k^{\flat} \xi_I} \right) \rangle
\end{align}
in terms of Weyl spinors. Repeated R-symmetry indices in a product are summed. Note that if the supersymmetry variation is picked to be aligned along the Dirac spinors associated to the spin quantization axis through the massless momentum $q$ the form of the supersymmetry variation is exactly the same as in the massless case. Also, the massless limit is easily recognized in the above.

Sums over the states of the super-multiplet which will appear in cuts below can be written as fermionic integrations over the coherent state parameters. This simply follows from expanding the amplitudes and performing the integration.
\begin{equation}
\sum_{s \in \textrm{multiplet}} A(\ldots, \{s, l_{\mu}\}) A(\{s, -l_{\mu}\}, \ldots) \rightarrow \int (dF)  A(\ldots, \{\eta,\iota, l_{\mu}\}) A(\{\eta,\iota, -l_{\mu}\}, \ldots)
\end{equation}
Here the shorthand
\begin{equation}\label{eq:defoffermmeasure}
(dF) = d\eta_{1} d\iota_{1} d\eta_{2} d\iota_{2}
\end{equation}
was used. There is an analogous formula for the conjugate coherent state, featuring the conjugate fermionic measure
\begin{equation}\label{eq:defoffermmeasureconj}
(\overline{dF}) = d\overline{\eta}_{1} d\overline{\iota}_{1} d\overline{\eta}_{2} d\overline{\iota}_{2}
\end{equation}

The massive spinor helicity spinors which belong to the momentum $-l$ can be obtained from the ones for $l$ by a phase rotation as
\begin{equation}
l^{\flat}_{\alpha} \rightarrow - l^{\flat}_{\alpha} \qquad \textbf{or} \qquad l^{\flat}_{\dalpha} \rightarrow - l^{\flat}_{\dalpha}
\end{equation}
as can easily be checked from \eqref{eq:decomponemom}. For the purposes of this article there are two crucial points to take from the analysis:
\begin{itemize}
\item there exists a superspace which generalizes the massless case smoothly.
\item the extra terms in the supersymmetry transformations of the coherent states are proportional to $\sim \frac{\braket{\chi \xi}}{\braket{k^{\flat} \xi}}$ and $\sim \frac{\sbraket{\chi \xi}}{\sbraket{k^{\flat} \xi}}$ and contain the opposite chirality spinors.
\end{itemize}
These features can be guessed based on knowledge of the fundamental massive multiplet analysis in \cite{Schwinn:2006ca} for $\mathcal{N}=1$, the realization that it is all just representation theory of the on-shell supersymmetry algebra as in \cite{Grisaru:1977px} and the fact that the $\mathcal{N}=4$ BPS representation is isomorphic to the fundamental massive $\mathcal{N}=2$ multiplet \cite{Fayet:1978ig}.

\subsection{Isolating triangle coefficients at one loop}
In general an amplitude with massive particles can at one loop be expanded into a sum over basis functions. The reduction argument is formed by repeated application of Passarino-Veltman reduction, combined with a reduction formula for scalar integrals. For a review of this argument and references to the original literature see the introduction of \cite{Badger:2008cm}. The upshot is that any amplitude can be expanded into a basis of scalar functions,
\begin{equation}\label{eq:massiveloopampexp}
A^{1-\textrm{loop}} = \sum a_b \left(\textrm{Boxes} \right) +  a_t  \left(\textrm{Triangles} \right) + a_{bb}  \left(\textrm{Bubbles} \right) + a_{tp}  \left(\textrm{Tadpoles} \right) + \textrm{Rational}
\end{equation}
with certain coefficients which are in general rational functions of momentum, multi-linear in the polarization vectors. Dimensional regularization in the four dimensional helicity scheme \cite{Bern:1991aq} is understood in this expression.   In this article the bubble and tadpole integrals are irrelevant as they are UV divergent and are therefore expected to be absent in $\mathcal{N}=4$ at any point of the moduli space.

The coefficients for the box function are easily isolated by a quadruple cut \cite{Britto:2004nc}. A triple cut will typically mix box and triangle contributions. If the box coefficients are known however, the triangle ones can be calculated from this triple cut. As first realized by Forde \cite{Forde:2007mi} in the massless case, the triangle and lower coefficients can also be isolated more directly. After a triple cut for instance the loop integral reduces to a single integral. The integrand, considered as a function of the complexified remaining parameter, has poles at finite distances in the complex plane as well as a pole at infinity. The finite distance poles are connected to the box coefficients as they are associated to an extra propagator going on-shell, while the pole at infinity isolates the triangle coefficient. This argument was extended to the massive case by Kilgore \cite{Kilgore:2007qr}.

\FIGURE[!h]{\epsfig{file=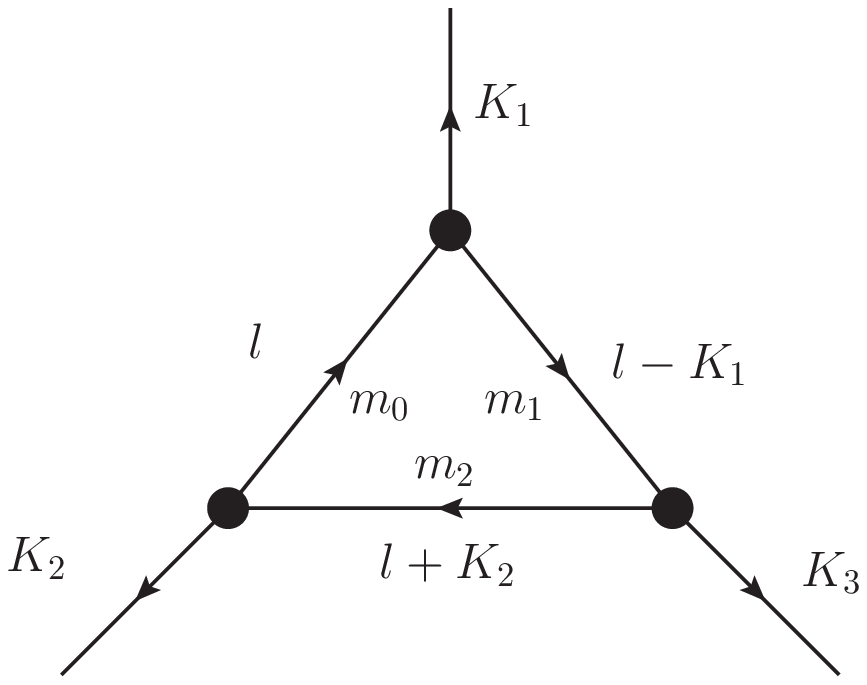,scale=0.9} \caption{Triple cut conventions} \label{fig:tripcutconv}}

\subsubsection*{Triangle coefficients}
In brief, at a triple cut with conventions as in figure \ref{fig:tripcutconv} we must have,
\begin{equation}
l^2 = m_0^2 \quad (l-K_1)^2 = m_1^2 \quad (l+K_2)^2 = m_2^2
\end{equation}
Using $K_1$ and $K_2$ as defining momenta in \eqref{eq:massivemomentumexp} these constraints can be solved explicitly by
\begin{equation}\label{eq:solloopmom}
l_{\mu} = \alpha_1 a^1_{\mu} + \alpha_2 a^2_{\mu}  + \frac{\alpha_1 \alpha_2 - \frac{m_0^2}{\gamma_{12}}}{t} a^3_{\mu}  + t a^4_{\mu}
\end{equation}
with
\begin{align}
 \alpha_1 & = \frac{K^2_2 K^2_1 -(m^2_2+m^2_0) \gamma_{12}- (m^2_0 + m^2_1+\gamma_{12}) K^2_1 }{K^2_1 K^2_2 - \gamma_{12}^2 }\\
 \alpha_2 & = \frac{- K^2_2 K^2_1 +(m^2_1+m^2_0) \gamma_{12} + (m^2_2 + m^2_0-\gamma_{12}) K^2_2 }{K^2_1 K^2_2 - \gamma_{12}^2 }
\end{align}
For future purposes it is useful to note that
\begin{equation}
\left(\begin{array}{ccc} l_{\mu} & = & l^{\flat}_{\mu} + \frac{m_0^2}{2 (l\cdot a^{3})} a^3_{\mu}\\
l^{\mu}-K^{\mu}_1 & = & (l-k_1)^{\flat}_{\mu} + \frac{m_1^2}{2 (l-k_1)\cdot a^{3} } a^3_{\mu} \\
l^{\mu}+K^{\mu}_2 & = & (l+k_2)^{\flat}_{\mu} + \frac{m_2^2}{2 (l+k_2)\cdot a^{3} } a^3_{\mu}
\end{array} \right.
\end{equation}
holds for this particular parametrization with $t$-dependent \emph{massless} momenta defined by
\begin{equation}\left(
\begin{array}{cc}
l^{\flat}_{\mu} & = \left(\alpha_1 a^1_{\mu} + \alpha_2 a^2_{\mu}  + \frac{\alpha_1 \alpha_2 }{t} a^3_{\mu}  + t a^4_{\mu}\right) \\
(l-K_1)^{\flat}_{\mu} & =  \left(\left(\alpha_1-1\right) a^1_{\mu} + \left(\alpha_2 - \frac{K^2_2}{\gamma_{12}}\right)a^2_{\mu}  + \frac{\left(\alpha_1-1\right)\left(\alpha_2 - \frac{K^2_2}{\gamma_{12}}\right)  }{t} a^3_{\mu}  + t a^4_{\mu} \right)\\
(l+K_2)^{\flat}_{\mu} & = \left(\left(\alpha_1+\frac{K^2_1}{\gamma_{12}} \right) a^1_{\mu} + \left(\alpha_2+1\right) a^2_{\mu}  +\frac{\left(\alpha_1+\frac{K^2_1}{\gamma_{12}}\right)\left(\alpha_2 + 1 \right)  }{t} a^3_{\mu}  + t a^4_{\mu} \right)
\end{array}\right. \label{eq:masslessmomchan}
\end{equation}
Taking into account the above analysis of massive spinor helicity, the emergence of a special lightcone momentum $a^3$ suggests that this vector should play the role of the vector `$q$'. In other words, it is natural to take a basis of states in the cuts using $a^3$ to define the spin polarization axis.

On a certain triple cut the amplitude can be written as sums of tree amplitudes over the Lorentz quantum numbers in the intermediate channels by unitarity,
\begin{multline}
A^{1}|_{\textrm{triple}} = \int dt J_t \left( \sum_{s_1,s_2,s_3} A(\{s_1, -l\}, \{s_2, l-K_1\}, X_1) A(\{s_2, -l+K_1\}, \{s_3, l+
K_2\}, X_2) \right. \\ \left. \phantom{\sum_{s_1}} A(\{s_3, -l-K_2\}, \{s_1, l\}, X_3) \right)
\end{multline}
Here the loop momentum is parameterized as in equation \eqref{eq:solloopmom} and $J_t$ is the Jacobian in the transformation of the integral to the above form. In this expression $X_1$, $X_2$ and $X_3$ stand for the momenta and Lorentz quantum numbers of the external states. This gives the left-hand side of the generic expansion \eqref{eq:massiveloopampexp}. The triangle coefficients $a_t$ from the generic expansion \eqref{eq:massiveloopampexp} can be isolated from the integrand of this expression \cite{Forde:2007mi}, \cite{Kilgore:2007qr} as
\begin{multline}\label{eq:isolatetrianglecoefs}
a_t = \left( \lim_{t \rightarrow \infty} \sum_{s_1,s_2,s_3} A(\{s_1, -l\}, \{s_2, l-K_1\}, X_1) A(\{s_2, -l+K_1\}, \{s_3, l+
K_2\}, X_2) \right. \\ \left. \phantom{\sum_{s_1}} A(\{s_3, -l-K_2\}, \{s_1, l\}, X_3) \right)_{t=0}
\end{multline}
In short, the instruction is to obtain a particular coefficient the Laurent expansion around $t=\infty$ of the particular triple product of amplitudes. It will be shown below that the expression on the right hand side of this equation vanishes  for maximally supersymmetric theories in four dimensions for every possible cut thereby proving absence of triangles. For this to work properly one more ingredient is needed, presented in the next section.

\section{Relation between broken and higher dimensional gauge theories}
\label{sec:stringpov}
In this section a map is derived between Lagrangians in higher dimensional gauge theory and lower dimensional gauge theory on the Coulomb branch. This map is motivated from consideration of a system of $N_c$ parallel $D_p$-branes separated along a line in general position where the dimension of the world-volume is taken to be strictly less than the ambient space dimension. This is of independent interest and will be presented first. Readers only interested in field theory may wish to skip directly to section \ref{sec:fieldtheorypov}.

\FIGURE[!h]{\epsfig{file=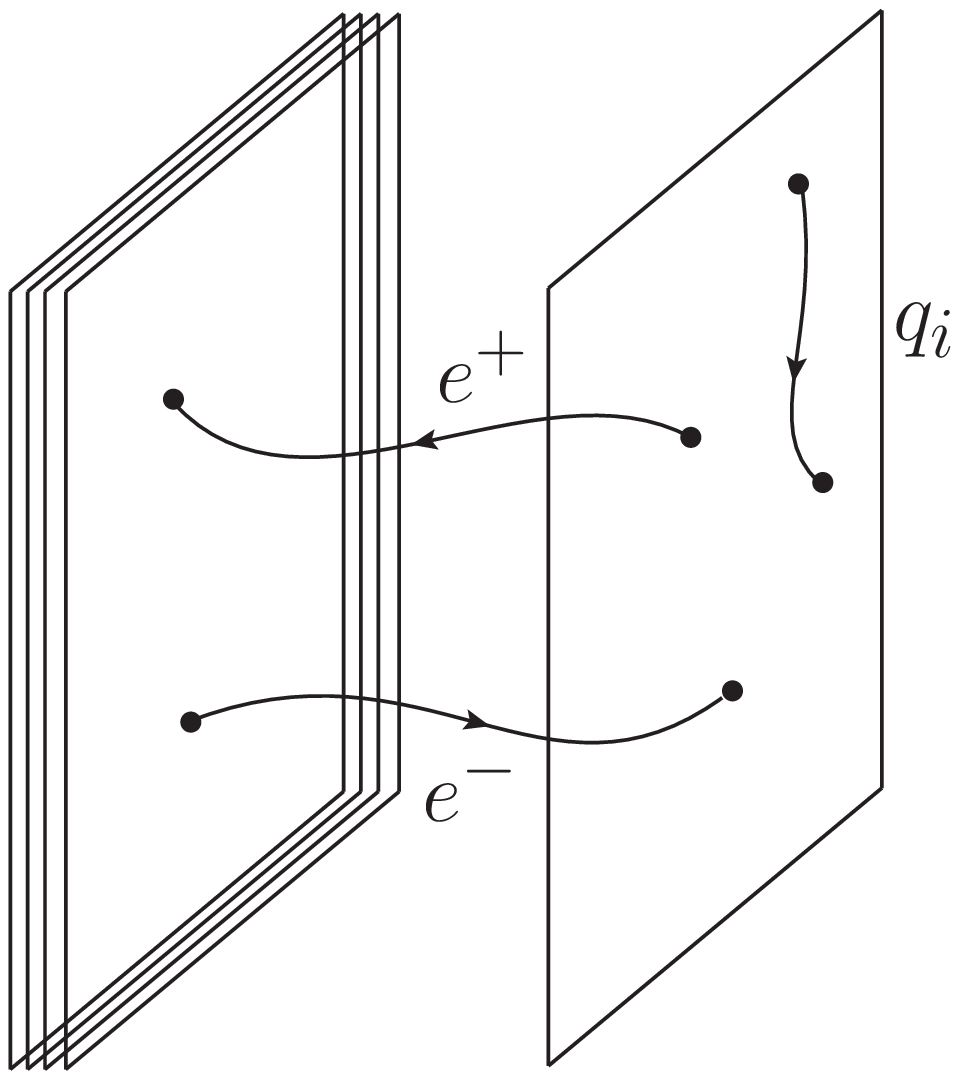,scale=0.6} \caption{Strings stretching between branes and associated Chevalley generators for the vertex operators} \label{fig:stringsbranes}}

\subsection{Color ordering in the broken phase from the D-brane picture}
The open string system can be interpreted from the worldsheet point of view as a system with $N_c$ degenerate vacua, each labeled by the position of the corresponding D-brane in the perpendicular direction. A string stretching between two branes changes the (perpendicular) position of the string. What is needed is a way of keeping track of how the string stretches between branes. This is provided by the usual color ordering prescription in case the branes are all coincident. Below this is presented in a way which makes clear that with a simple modification the same prescription can be used to study non-coincident branes.

\subsubsection*{Two branes}
Let the two vacua be given by states
\begin{eqnarray}
|0\rangle_1 \leftrightarrow \left(\begin{array}{c} 1 \\ 0 \end{array}\right) & \quad &
|0\rangle_2 \leftrightarrow \left(\begin{array}{c} 0 \\ 1 \end{array}\right)
\end{eqnarray}
which are clear eigenvectors of both
\begin{eqnarray}
h = \left(\begin{array}{cc} 1 & 0 \\ 0 & -1 \end{array}\right) & \textrm{and} & i = \left(\begin{array}{cc} 1 & 0 \\ 0 & 1 \end{array}\right)
\end{eqnarray}
It is clear one state can be transformed into the other by
\begin{eqnarray}
e^+ = \left(\begin{array}{cc} 0 & 1 \\ 0 & 0 \end{array}\right)  & \quad &
e^- = \left(\begin{array}{cc} 0 & 0 \\ 1 & 0 \end{array}\right)
\end{eqnarray}
Note in particular that
\begin{equation}
e^+ |0\rangle_1 = e^- |0\rangle_2 = 0
\end{equation}
which simply reflects the physical intuition that a string can only stretch from brane $1$ to brane $2$ if it is attached to brane $1$ to begin with. It is easy to recognize in $e^+,e^-,h$ the standard Chevalley basis generators of the Lie algebra $A_1$ one of whose real forms is better known as $SU(2)$.

A scattering amplitude will in general correspond to the vacuum to vacuum transition probability,
\begin{equation}
A \sim \sum \phantom{p}_i\langle 0|V_1 \ldots V_n  |0\rangle_i
\end{equation}
where a sum over the two degenerate vacua and a generic sequence of creation operators is inserted. To keep track of the stretching of the string every vertex operator which creates a mode between branes will now be associated with a $e^+$ or $e^-$ generator depending on orientation, while strings stretching from brane $1$ to $1$ or $2$ to $2$ are naturally associated with the generators
\begin{equation}\label{eq:measuringposbranestwo}
q_1 = \frac{1}{2} \left(h+i \right) \quad \quad q_2 = \frac{1}{2} \left(h-i \right)
\end{equation}
respectively. Hence every string amplitude at tree level is naturally associated with a sum over group theory traces,
\begin{equation}\label{eq:colorordgen}
\tr \left(T^{(1)} \ldots T^{(n)}\right)
\end{equation}
where the ordering depends on the ordering of the particles in the amplitude. From
\begin{equation}
(e^+)^2 = (e^-)^2 = 0
\end{equation}
it is clear that in any given color trace many terms will generically vanish. It is easy to check that this happens in instances for which there is no consistent way of labeling the boundaries of the string worldsheet. Note that the mass of the particles in the ordered amplitude are correlated with the group theory factors. The resulting group theory trace is of course the usual `Chan-Paton' factor \cite{Paton:1969je}, but now in a preferred basis of generators.

As the above analysis only keeps track of the branes on which the string ends and not on the positions of these branes, it follows that also in the separated brane case the same color ordering factor will be obtained. For coincident branes the basis of generators can be changed at will as the branes are indistinguishable. For separated branes that is no longer the case. Note that for unbroken theories it might also be interesting to study (field and string theory) amplitudes in this particular basis. In other cases such as Einstein gravity \cite{BjerrumBohr:2008ji} or QED \cite{Badger:2008rn} it is known that the involved permutation sums can drastically simplify resulting expressions.

Of course, the above analysis is easily generalized to a system of $N_c$ branes. One can take a simple basis of raising
\begin{equation}\label{eq:basisofUN1}
\left(e_{km}^+\right)_{ij} = \delta_{i,k} \delta_{j,m} \qquad k<m
\end{equation}
for strings stretching from brane $k$ to $m$ and similar lowering operators,
\begin{equation}\label{eq:basisofUN2}
\left(e_{km}^-\right)_{ij} = \delta_{i,k} \delta_{j,m} \qquad k>m
\end{equation}
which represent strings stretching from $m$ to $k$ in addition to the commuting set of generators
\begin{equation}\label{eq:basisofUN3}
\left(q_k\right)_{ij} = \delta_{i,k} \delta_{j,k}
\end{equation}
This is a Chevalley-like basis of generators of the defining representation of $U(N_c)$. The identity $U(1)$ generator together with the $SU(N_c)$ Cartan sub-algebra forms an Abelian sub-algebra which measures on which brane the string lives, while the $SU(N_c)$ root generators model strings stretching from one brane to an adjacent one. The trace structure for the scattering amplitudes is the obvious generalization of the Chan-Paton factor of equation \eqref{eq:colorordgen}.

\subsection{Open string vertex operators for massive vector bosons}
The vertex operators for the modes of the bosonic string stretching between branes have been discussed in \cite{Pesando:1999hm}. That these exist follows from the operator-state correspondence. Here a different approach is followed which stresses the connection to the field theory. Consider the calculation of a scattering amplitude on a stack of parallel branes in the string theory from the path integral point of view. For a theory without vertex operators a background can be incorporated into the action by including the usual background field coupling,
\begin{equation}\label{eq:backgroundfieldcoupling}
P e^{\oint A^{\mu}(X) d X_{\mu} (+ F_{\mu\nu} \psi^{\mu} \psi^\nu)}
\end{equation}
Here the term between brackets is included in the RNS superstring. The integral is along the boundary of the worldsheet. With vertex operators this would simply be a Wilson line stretching from one vertex operator to the next, representing a coherent background of fields. This term needs to be considered for a constant background gauge field valued in the Cartan sub-algebra, pointed along a direction perpendicular to the brane, say
\begin{equation}
\braket{A_{\mu}} = n_{\mu} \left( \sum_i c_i  q^i\right)
\end{equation}
with $q_i$ the commuting set of generators defined above which is closely related to the Cartan subalgebra. For a disc the term \eqref{eq:backgroundfieldcoupling} seems to vanish identically in this background. However, for a scattering amplitude with $n$ particles, the topology to consider is that of a disc with $n$ marked points. Hence for every marked point, the corresponding vertex operator picks up a gauge transformation factor,
\begin{equation}
V_a(z_1) T^a \rightarrow e^{n_{\mu} \left( \sum_i c_i  q^i\right) X^{\mu}(z_1)}\left( V_a(z_1) T^a \right) e^{-n_{\mu} \left( \sum_i c_i  q^i\right) X^{\mu}(z_1)}
\end{equation}
The background field is perpendicular to the brane or more precise, to the momentum of the excitation of the open string on the brane. This makes the multiplication in this expression well-defined from the CFT point of view. It is easy to see that it is natural to take a Chevalley basis as above for the generators of the vertex operator, $T$. In this case,
\begin{multline}
e^{n_{\mu} \left( \sum_i c_i  q^i\right) X^{\mu}(z_1)} \left( V_i(z_1) T^i \right)  e^{-n_{\mu} \left( \sum_i c_i  q^i\right) X^{\mu}(z_1)} = \\ \left\{\begin{array}{cccc} \left( V_i(z_1) T^i \right) e^{ \left(c_k - c_m\right) n_{\mu} X^{\mu} (z_1)}&  \qquad  & \textrm{if} & T = e_{km}^+ \\
\left( V_i(z_1) T^i \right) e^{\left(c_m - c_k\right) n_{\mu} X^{\mu} (z_1)} &  \qquad & \textrm{if} & T = e_{km}^-
\\ \left( V_i(z_1) T^i \right) &  \qquad &  & \textrm{else} \end{array} \right.
\end{multline}
Here no summation on the index $i$ is implied. In other words, the vacuum expectation value of the background field translates into a 'momentum' proportional to $v_i n^{\mu}$ in the direction perpendicular to the two branes. The vertex operators for the non-Cartan subalgebra elements of the Lie algebra are therefore very closely related to the open string vertex operators on a higher dimensional brane, where the vev plays the role of the higher dimensional momentum. This is exactly the result found by Pesando \cite{Pesando:1999hm}. Note that higher dimensional momentum conservation neatly translates into the fact that there should be an equal number of $e^+_{km}$ and $e^-_{km}$ generators to generate a non-zero Chan-Paton factor.

From the point of view of the conformal field theory, it is obvious that the obtained vertex operators have the right commutation relations with the modes of the energy momentum tensor: this is a simple consequence of their higher dimensional interpretation. The quantum numbers follow from the usual free field analysis. In particular, the mass-shell condition as obtained by the constraint equations of the Virasoro algebra become
\begin{equation}
k^2 = (c_k - c_m)^2
\end{equation}
for the modes of the string associated to the $e^+_{km}$ and $e^-_{km}$ generators where $k$ is the momentum along the stack of branes, while the polarization vector $\xi$ obeys
\begin{equation}
p^{\mu} \xi_{\mu} = 0 \qquad \xi_{\mu} \sim \xi_{\mu} + g p_{\mu}
\end{equation}
with $p$ the higher dimensional momentum. The spectrum of the string theory therefore changes as indicated above. Just as above, the massless representations of the supersymmetry algebra in higher dimensions become BPS representations in the lower dimensions.

The derivation of scattering amplitudes now proceeds along the same lines as before. In order to have conformal invariant scattering amplitudes vertex operators integrated along the boundaries should be used - after having fixed three to fixed positions due to residual conformal invariance. Due to the integration along the boundaries, the scattering amplitude will involve a sum over non-cyclicly connected different orderings of the vertices, each of which involves its own Chan-Paton factor. Hence a color ordering for the broken symmetry case is obtained as derived above from general arguments. Again, the difference to the usual analysis for coincident branes is that the color traces have to be calculated in the Chevalley basis.

The map between modes of stretching strings and higher dimensional vertex operators works fine for  tachyons and massless gluons in the open bosonic string, but seems tricky in the superstring. The problem is that only an exponential is picked up by the above reasoning, while the vertex operator reads
\begin{equation}
\xi_{\mu} \left(\partial X^{\mu} + \psi^{\mu} \psi^{\nu} k_{\nu} \right) e^{\ii k X}
\end{equation}
Hence there is a discrepancy between massive vector and massless higher dimensional vector vertex operators as obtained by the above reasoning. It would be very interesting to explore this further, but for the purposes of this article the field theory limit is enough. The discrepancy will be sub-leading in this limit. The author suspects, but does not prove, that the naive map between higher dimensional vertex operators and stretching modes of the string also holds in the superstring.

\subsection*{Some applications in string theory}
Several results follow immediately from the above string theory picture. If the conjecture about the superstring is true for instance, the six point scattering amplitudes in the spontaneous symmetry broken case can for instance be constructed order by order in $\alpha'$ from the higher dimensional result in the unbroken case in the literature \cite{Oprisa:2005wu}. Some more results follow:

\subsubsection*{Goldstone boson equivalence theorem in string theory}
The gauge symmetry of the higher dimensional string theory,
\begin{equation}\label{eq:highdimgaug}
e_{\mu} \rightarrow e_{\mu} + f k_{\mu}
\end{equation}
for some function f which follows from decoupling of zero norm states can be used to derive the string theory equivalent of the field theory Goldstone boson equivalence theorem (see for instance \cite{Peskin:1995ev} and references therein). This theorem follows from the kinematical relation between the four dimensional momentum and the longitudinal polarization of a massive vector boson,
\begin{equation}
k^{\mu} = m \epsilon^{\mu}_0 + \frac{m^2}{q\cdot k} q^{\mu}
\end{equation}
Using this together with the (higher dimensional) gauge symmetry of \eqref{eq:highdimgaug} shows in a covariant way that the longitudinal polarization is related to the scalar mode which gets a vev. Working this relation out in a fixed Lorentz frame reproduces the analysis in \cite{Peskin:1995ev}.

\subsubsection*{On-shell recursion}
By the relation to the higher dimensional gauge theory it is easy to see that some form of on-shell recursion relations will hold for amplitudes in the spontaneously symmetry broken theory. In the higher dimensional theory recursion has been proven in Yang-Mills theory in \cite{ArkaniHamed:2008yf} and in string theory (nearly) simultaneously in \cite{Cheung:2010vn} and \cite{Boels:2010bv} after a conjecture in \cite{Boels:2008fc}. The string theory derivation immediately applies to the broken symmetry phase by the analysis above, at the least in the bosonic string. However, shifts in the higher dimensional theory will when interpreted in the lower dimensional broken theory generically shift the momenta in the directions perpendicular to the brane. The terms in the recursion relation therefore will for a generic shift therefore involve amplitudes with different symmetry breaking vevs. The particular shift studied above for massive vector bosons does stay within the class of amplitudes with a fixed vev.

\begin{table}
\begin{center}
\begin{tabular}{c|c c c}
$e_i \;\backslash \;e_j $ &  $-$  &  $+$  & T \\
\hline
$-$                           & $ +1$ & $ +1$ & $ +1$ \\
$+$                           & $ -3$ & $ +1$ & $ -1$ \\
T                             & $ -1$ & $ +1$ & $ -1$ \\
T2                            & $ -1$ & $ +1$ & $  0$ \\
\end{tabular}
\caption{upper bounds on the leading power in $z^{-\kappa}$ in the large $z$ limit for adjacent shifts of two shifted polarized massive vector bosons in general $\mathcal{N}=2$ gauge theories on the Coulomb branch.}
  \label{tab:BCFWkinmasveccoulomb}
\end{center}
\end{table}

As in string theory BCFW shifts for higher dimensional theories have a smooth field theory limit, on-shell recursion relations also hold for amplitudes in the broken field theory. In particular, in comparison with the naive Feynman diagram power counting of Table \ref{tab:BCFWkinmasvec} the behavior of the $(++)$ and $(--)$ color adjacent shift is improved for the particular class of symmetry breaking theories under study. This yields Table \ref{tab:BCFWkinmasveccoulomb} from the higher dimensional interpretation in the field theory limit ($\al \rightarrow 0$). Before the limit there is only a universal prefactor
\begin{equation}
\sim z^{\al (k_1 + k_n)^2}
\end{equation}
for adjacent shifts of legs one and $n$ to take into account in the superstring case (see \cite{Cheung:2010vn}, \cite{Boels:2010bv} for the specifics in the bosonic string case). For non-adjacent shifts the analysis from \cite{Boels:2010bv} also applies to the case at hand. The transverse modes translate directly into specific combinations of scalars and vector fields using the explicit polarization states of \cite{Boels:2009bv}. The difference between $T$ and $T2$ is wether the transverse polarizations on the different legs have a vanishing inner product or not.

\subsubsection*{Monodromy relations}
The co-cycle factor between two vertex operators in any sector of the theory is unchanged with the convention that the D-brane separation is included as 'extra dimensional momentum'. Hence
\begin{equation}
:\!V_1(z_1)\!:\, :\!V_2(z_2)\!:  = :\!V_2(z_2)\!:\, :\!V_1(z_1)\!:\, e^{2 \al k_1 k_2 \frac{(z_1-z_2)}{|z_1 - z_2|}}
\end{equation}
This can be used to show as expected that the color ordered amplitudes remain cyclic. Moreover, this can be used to derive relations between different color-ordered amplitudes as shown in \cite{Boels:2010bv}, just as in \cite{Plahte:1970wy} (see also \cite{BjerrumBohr:2009rd}). This was also used recently to derive a minimal basis of amplitudes in \cite{BjerrumBohr:2009rd}. The only extra ingredient is that now track should be kept of the mass of the different particles involved (or equivalently of their explicit color factor). These relations drive the non-adjacent shift analysis for on-shell recursion relations.

\subsection{Generalized dimensional reduction in field theory}
\label{sec:fieldtheorypov}
Although above a string theory viewpoint was presented, in the field theory limit
\begin{equation}
\al \rightarrow 0 \quad \quad \sqrt{\al} (c_k - c_m) \,\, \textrm{fixed}
\end{equation}
the usual spontaneously symmetry broken theory is obtained. Moreover, the string theory picture presented above should be derivable directly in the field theory and this will be done in this subsection.  In field theory a similar construction has been used to study a particular class of scattering amplitudes in \cite{Selivanov:1999ie}.

The key insight is that a class of spontaneously symmetry broken Lagrangians can be obtained by a form of generalized dimensional reduction from a higher dimensional massless theory. This follows by the observation that for a vector $n_i$ pointing in a direction perpendicular to the dimensions of interest here the covariant derivative reads
\begin{equation}\label{eq:covariantderivarg}
n_i^\mu \left(\partial_{\mu} + A_{\mu}\right) = \left(n_i^\mu \partial_{\mu} + \phi\right)
\end{equation}
with $n_{i}^{\mu} A_{\mu} \equiv \phi$ one of the lower dimensional scalar modes. In ordinary dimensional reduction the off-dimensional momenta is set to zero. The structure of the derivative shows that giving this scalar mode a vev,
\begin{equation}
\phi \rightarrow \tilde{\phi} + \braket{\phi}
\end{equation}
can be traded for a higher dimensional momentum. To make this precise, consider for instance a minimally coupled matter scalar $\psi$ of the higher dimensional theory valued in the adjoint and expand it in the basis of $U(N)$ adjoint generators of equations \eqref{eq:basisofUN1}, \eqref{eq:basisofUN1} and \eqref{eq:basisofUN3} as
\begin{equation}
\psi = \sum_{k<m} \psi^{+,km} e^{+}_{km} +  \sum_{k>m} \psi^{-,km} e^{-}_{km}  + \sum_{k} \psi^{0,k} e^{0}_{k}
\end{equation}
For a scalar vev
\begin{equation}
\braket{\phi} = \sum c_k e^{0}_{k}
\end{equation}
one obtains
\begin{equation}
\left([\braket{\phi}, \psi]\right) = \sum_{k<m} (c_k - c_m) \psi^{+,km} e^{+}_{km} -  \sum_{k>m} (c_k - c_m)  \psi^{-,km} e^{-}_{km}
\end{equation}
To obtain the Feynman rules the Lagrangian has to be expanded around the scalar vev. The same expansion is obtained on the level of the Lagrangian if the scalar field has no vev, but the component fields $\psi^{+,km}$ and $\psi^{-,km}$ acquire a constant 'extra-dimensional' momentum as
\begin{equation}\label{eq:momassignadj}
n^{\mu} \partial_{\mu} \psi^{+,km} = - n^{\mu} \partial_{\mu} \psi^{-,km} = c_k - c_m
\end{equation}
while the other components ($\psi^{0,k}$) are left massless. The same argument applies to minimally coupled fermionic matter.

There is also a natural basis for minimally coupled matter valued in the fundamental given by
\begin{equation}
\psi = \sum_i \psi_i \tilde{e}^i
\end{equation}
with
\begin{equation}
e^{0}_{k} \tilde{e}^i = \delta_k^i \tilde{e}^i
\end{equation}
Here the equivalent expansion involves momenta
\begin{equation}
n^{\mu} \partial_{\mu} \psi_i =  c_i
\end{equation}
which follows trivially from the covariant derivative.

A similar argument also applies to the relation between lower and higher dimensional gauge fields. For this one uses,
\begin{equation}
F_{\mu\nu} \sim [D_{\mu},D_{\nu}]
\end{equation}
In other words, the gauge field gets a constant off-dimensional momentum like equation \eqref{eq:momassignadj} just as adjoint matter. It is instructive to make this explicit in an example relevant to the theory of main interest in this article. The $\mathcal{N}=2$ gauge multiplet part of the $\mathcal{N}=4$ Lagrangian in $D=4$ is that describing the dimensional reduction of the six-dimensional Yang-Mills Lagrangian
\begin{equation}
\mathcal{L}^{D=4}_{\textrm{rel}}  = \tr \left( D_{\mu} \phi D_{\mu} \overline{\phi}  + [\phi, \overline{\phi}]^2 + (F_4)^2\right)
\end{equation}
here the superscript on the scalar field has been dropped for notational convenience. This term arises in the dimensional reduction from the action of a six-dimensional Yang-Mills theory.
\begin{equation}
\mathcal{L}^{D=4}_{\textrm{rel}} = \tr \left(F_{6,\mu\nu} F_{6,\mu\nu} \right)_{\textrm{red}}
\end{equation}
In more detail, the six dimensional field strength $F_6$ can be reduced w.r.t. a chosen four dimensions as
\begin{equation}
F_6 = \left(\begin{array}{c|cc} F^4_{\mu\nu} & D_{\mu} A_5 & D_{\mu} A_6   \\
\hline
\cdots & 0 &  [A_5,A_6] \\
\cdots & \cdots & 0 \\
\end{array} \right)
\end{equation}
where in particular the momenta in the $5,6$ directions have been set to zero. The symbol `$\cdots$' indicates non-trivial parts of the matrix determined by its antisymmetry. The Lagrangian in the symmetry broken phase in four dimensions can be obtained from the Lorentz trace ($(F_6)_{\mu\nu} (F_6)^{\mu\nu}$) of this expression by expanding around the relevant vev, taken to be
\begin{equation}
\langle A_5 + \ii A_6 \rangle = \sum_i (c_5 + \ii c_6)^i  q_i
\end{equation}
In terms of the new fields $\tilde{A}_5, \tilde{A}_6$ obtained by
\begin{equation}
A_5 + \ii A_6 \rightarrow \left( \sum_i (c_5 + \ii c_6)^i  q_i \right) + \tilde{A}_5 + \ii \tilde{A}_6
\end{equation}
the six dimensional field strength in these new field coordinates reads
\begin{equation}\label{eq:compFtosixdim}
F_6 = \left(\begin{array}{c|cc} F^4_{\mu\nu} & D_{\mu} \tilde{A}_5 + \sum_i c^i_5 [A_{\mu}, q_i] & D_{\mu} \tilde{A}_6 + \sum_i c^i_6 [A_{\mu}, q_i]  \\
\hline
\cdots & 0 &  [\tilde{A}_5,\tilde{A}_6] + \sum_i c^i_5 [q_i,\tilde{A}_6] - \sum_i c^i_6 [q_i,\tilde{A}_5] \\
\cdots & \cdots & 0 \\
\end{array} \right)
\end{equation}
where $A_{\mu}$ is the four dimensional gauge field. Hence with the momentum assigned the same as for the adjoint scalar in equation \eqref{eq:momassignadj} it is seen that the higher dimensional and the broken gauge theory are simply related.

Note this can easily be extended to $\mathcal{N}=4$ in $D=4$ as this can be considered to be a six dimensional gauge theory coupled to adjoint matter. The above argument is quite general and applies for instance to the full moduli space of $\mathcal{N}=4$ in $D=4$, not just the part of it parameterized by one complex scalar. Moreover, it encompasses the full Coulomb branch of $\mathcal{N}=2$ in $D=4$ coupled to fundamental matter. Note that the higher dimensional momenta immediately translate into lower dimensional BPS charges by the just derived map if supersymmetric theories are considered.

\subsection*{Perturbation theory}
Since the Lagrangians of the higher dimensional and broken gauge theory are equivalent under the above map, the perturbation theory at least at the tree level must be equivalent. The gauge fixing can be related by the same map as above in both cases. Note that for instance the higher dimensional Feynman-'t Hooft gauge for $\xi=1$ reduces in particularly nice way. The color ordering prescription also descends from the higher dimensional theory, exactly as in the discussion of color ordering in the string theory.

Concretely, to calculate a tree level amplitude in the broken gauge theory one takes a higher dimensional amplitude, fixes a Chevalley basis for the generators and restricts the momenta in the extra dimensions as in the setup above for all states which become massive in four dimensions (i.e. for the full $A_{km}^{pm}$ $\mathcal{N}=4$ multiplet for the main example of this article). Higher dimensional momentum conservation becomes neatly entangled with the Chevalley basis quantum numbers. Since at tree level the external quantum numbers determine the internal ones, this finishes the argument: for every propagator in the higher dimensional theory which gets a mass, the external group quantum numbers match this. This can be seen by labeling the boundaries between external fields in a consistent way as if they ended on different branes. The polarizations in higher and four dimensions are related as explained in \cite{Boels:2009bv}. By the results of the latter article, a class of vanishing amplitudes is also immediately obtained in the four dimensional theory, which are the analogs of the helicity equal amplitudes in the massless case. Unfortunately, the analog of the one helicity opposite amplitude only vanishes for special choices of polarization axis.

There is an important subtlety in the perturbation theory related to the gauge coupling as these are related formally as
\begin{equation}
\frac{1}{g_4^2} \equiv \frac{1}{g_6^2} \int dx^2
\end{equation}
with the integration over the orthogonal space to the four dimensions of interest. At tree level this is of no concern, but at loop level there is a definite differences as one is instructed to integrate over momenta in different dimensions. In other words, the external quantum numbers do not determine the internal ones any more. However, there is a point on the moduli space of the higher dimensional integrand where the loop momentum can be chosen to be correlated with the gauge index of the internal loop. On this particular point this reduces the higher dimensional integrand to the lower dimensional spontaneously broken one.

Note that on-shell recursion for tree level amplitudes in the broken theory will work by the higher dimensional field theory derivation of \cite{ArkaniHamed:2008yf}. The BCFW shift behavior as a function of the spin of the shifted legs is captured in Table \ref{tab:BCFWkinmasveccoulomb}. The analysis of this subsection also shows clearly the close relation of the regularization proposed in \cite{Alday:2009zm} to dimensional regularization.

\section{No triangles on the moduli space of $\mathcal{N}=4$ in $D=4$}
\label{sec:notriangles}
The pieces of the puzzle introduced in the previous sections will now be assembled. First an on-shell proof of no-triangles will be presented, followed by an off-shell version which also proves `no rationals'.

\subsection{On-shell}
In the on-shell approach introduced in \cite{ArkaniHamed:2008gz} the first step is to rewrite the sums in \eqref{eq:isolatetrianglecoefs} as integrals over coherent states. Then supersymmetry is used to find an expression of the triangle coefficients in terms of sums over a large momentum limit of amplitudes with fixed spin quantum numbers for the internal legs. This reduces the problem to finding the large momentum limit of a restricted class of tree amplitudes.  Particular care is taken to trace the dependence on the parameter $t$ through all spinor expressions. The limiting behavior of the resulting expression can be analyzed directly and explicitly using the above relation to higher dimensional field theory.

To use the coherent state setup a choice must be made for the coherent state representation in each channel. This includes a choice of spin polarization axis and a choice of top state. Based on the above analysis of the triple cut and in particular equation \eqref{eq:masslessmomchan} it is natural to choose the same spin polarization axis for all cut channels generated by the light-like vector $a^3$,
\begin{equation}\nonumber
\left(a^{3}\right)^{\alpha \dalpha} = k_1^{\flat, \alpha} k_2^{\flat, \dalpha} \equiv  1^{\alpha} 2^{\dalpha}
\end{equation}
Furthermore for the coherent states spin up will be chosen to be the top state in channels $1$ (momentum labeled by $l$) and $2$ (momentum labeled by $l-K_1$) and spin down in the remaining one.

Using the coherent state representations for this choice of top states yields for the formula for the triangle coefficients
\begin{multline}
a_t = \lim_{t \rightarrow \infty} \int (dF)_1 (dF)_2 (\overline{dF})_3 A\left(\{\eta_1, \iota_1, l\}, \{\eta_2, \iota_2, l-k_1\}, X_1\right) \\ A(\{\eta_2, \iota_2, -(l-k_1)\}, \{\overline{\eta}_3, \overline{\iota}_3, l+k_2\}, X_2)  A(\{\overline{\eta}_3, \overline{\iota}_3, -(l+k_2)\}, \{\eta_1, \iota_1, - l\}, X_3)
\end{multline}
with the measures defined in \eqref{eq:defoffermmeasure} and \eqref{eq:defoffermmeasureconj}.  In this expression it is understood that $X_1$, $X_2$ and $X_3$ stand for the momenta and coherent state parameters of the external states of the particular triangle coefficient under study. In the following the notation
\begin{equation}
l^{\flat, \alpha \dalpha} = \lambda^{\alpha}_1 \lambda^{\dalpha}_1 \qquad (l-k_1)^{\flat, \alpha \dalpha} = \lambda^{\alpha}_2 \lambda^{\dalpha}_2 \qquad (l+k_2)^{\flat, \alpha \dalpha}  = \lambda^{\alpha}_3 \lambda^{\dalpha}_3
\end{equation}
will be used. Furthermore the choice of phases
\begin{equation}\label{eq:choiceofminussigns}
-l^{\flat, \alpha \dalpha} = \left(-\lambda^{\alpha}_1\right) \lambda^{\dalpha}_1 \qquad -(l-k_1)^{\flat, \alpha \dalpha} = \left(-\lambda^{\alpha}_2\right) \lambda^{\dalpha}_2 \qquad -(l+k_2)^{\flat, \alpha \dalpha}  = \lambda^{\alpha}_3 \left(-\lambda^{\dalpha}_3\right)
\end{equation}
will be made.

A further choice involves exactly how the massive spinor helicity spinors in the legs depend on $t$. This choice fixes the usual scaling ambiguity. From \eqref{eq:getspinorinnerprods} and \eqref{eq:masslessmomchan}
\begin{equation}
\braket{1 \lambda_i} \sbraket{2 \lambda_i} = t
\end{equation}
for instance follows which allows two natural choices. For both choices the other equations in \eqref{eq:getspinorinnerprods} can be solved to obtain either
\begin{equation}\label{eq:largetspinorsI}
\sbraket{1 \lambda_i} \sim \frac{1}{t} \quad \braket{1 \lambda_i} \sim t \quad \sbraket{2 \lambda_i} \sim 1 \quad  \braket{2 \lambda_i} \sim 1
\end{equation}
or
\begin{equation}\label{eq:largetspinorsII}
\braket{2 \lambda_i} \sim \frac{1}{t} \quad \sbraket{2 \lambda_i} \sim t \quad \braket{1 \lambda_i} \sim 1 \quad  \sbraket{1 \lambda_i} \sim 1
\end{equation}
where $i$ labels the cut legs. The choice will be made here that the first holds on leg $1$ and $2$, while the second holds on leg $3$.

There exists a finite supersymmetry transformation parameterized by Dirac spinors $\chi^I$, $\overline{\chi}^I$ which shifts $\eta_I$ and $\iota_I$ to zero. The transformations are given in terms of two $2$ component Weyl spinors of a complex $D=4$ Dirac spinor,
\begin{equation}
\left(\begin{array}{ll}
\chi^{I}_{\dalpha} & = \frac{\lambda_1^{\dalpha} \iota_2^I - \lambda_2^{\dalpha} \iota_1^I}{\sbraket{\lambda_1 \lambda_2}} \\
\chi^{I,\alpha}    & = 0
\end{array} \right.\qquad \qquad
\left(\begin{array}{ll}
\overline{\chi}^{I,\dalpha} & = \frac{\lambda_1^{\dalpha} \eta_2^I - \lambda_2^{\dalpha} \eta_1^I}{\sbraket{\lambda_1 \lambda_2}} \\
\overline{\chi}^I_{\alpha}  & = 0 \label{eq:chialph}
\end{array}\right.
\end{equation}
Since these two transformations do not commute, one should first perform one (say given by $\overline{\chi}$) and then the other. From the cut legs this picks up a four phase factors of the amplitudes on either side of cut $1$ and $2$ of the form
\begin{equation}\label{eq:phasefact}
\sim \left( e^{m_1 \iota_I \frac{\sbraket{\overline{\chi}^I 2}}{\sbraket{2 \lambda_1}}} \right)^2 \left( e^{m_2 \iota_I \frac{\sbraket{\overline{\chi}^I 2}}{\sbraket{2 \lambda_2}}} \right)^2
\end{equation}
Importantly, the only $t$ dependence in this phase is contained in the transformation parameter $\overline{\chi}$. Of course, the supersymmetry transformation also generates a transformation of the external states in $X_1$, $X_2$ and $X_3$ and of the third cut leg. For the external states the only dependence on $t$ is through both $\overline{\chi}$ and $\chi$. On the third cut leg the above supersymmetry transformation only shifts $\overline{\eta}_3$ and $\overline{\iota}_3$, since with the above convention for the phases in \eqref{eq:choiceofminussigns} the phase factors cancel on the two sides of the cut for the above finite supersymmetry transformation. Concretely, on one side
\begin{align}
\overline{\eta}_3 & \rightarrow \overline{\eta}_3 + m_3 \frac{\sbraket{\overline{\chi} 2}}{\sbraket{2 \lambda_3}}\\
\overline{\iota}_3 & \rightarrow \overline{\iota}_3 + m_3 \frac{\sbraket{2 \chi}}{\sbraket{2 \lambda_3}}
\end{align}
and on the other side
\begin{align}
\overline{\eta}_3 & \rightarrow \overline{\eta}_3 - m_3 \frac{\sbraket{\overline{\chi} 2}}{\sbraket{2 \lambda_3}}\\
\overline{\iota}_3 & \rightarrow \overline{\iota}_3 - m_3 \frac{\sbraket{2 \chi}}{\sbraket{2 \lambda_3}}
\end{align}
Leaving the terms proportional to $m_3$ for now, there exists a finite supersymmetry transformation which shifts $\overline{\eta}_3$ and $\overline{\iota}_3$ to zero,
\begin{equation}
\left(\begin{array}{ll}
\xi^{I}_{\dalpha} & = 0 \\
\xi^{I,\alpha}    & = - \frac{1^{\alpha}}{\braket{1 \lambda_3}} \overline{\iota}_3
\end{array} \right.\qquad \qquad
\left(\begin{array}{ll}
\overline{\xi}^{I,\dalpha} & = 0 \\
\overline{\xi}^I_{\alpha}  & = - \frac{1^{\alpha}}{\braket{1 \lambda_3}} \overline{\eta}_3
\end{array} \right.
\end{equation}
By the choice of phases in \eqref{eq:choiceofminussigns} for the cut legs $1$ and $2$ and the general structure of the supersymmetry transformation this transformation is seen not to generate any dependence on the momenta of those cut legs, while leaving the parameters of the coherent states untouched. Schematically, we have
\begin{multline}\label{eq:cohstatinttriangle}
a_t = \lim_{t \rightarrow \infty} \int \int (dF)_1 (dF)_2 (\overline{dF})_3 \,\,f(X_1, X_2, X_3) \,\,A\left(\{0,0, l\}, \{0,0, l-k_1\}, X_1\right) \\ A(\{0,0, -(l-k_1)\}, \{m_3 \frac{\sbraket{\overline{\chi} 2}}{\sbraket{2 \lambda_3}}, m_3 \frac{\sbraket{2 \chi}}{\sbraket{2 \lambda_3}}, l+k_2\}, X_2) \\ A(\{-m_3 \frac{\sbraket{\overline{\chi} 2}}{\sbraket{2 \lambda_3}}, -m_3 \frac{\sbraket{2 \chi}}{\sbraket{2 \lambda_3}}, -(l+k_2)\}, \{0,0, - l\}, X_3)
\end{multline}
where $f(X_1, X_2, X_3)$ contains the phases acquired from the susy transformations on the coherent state parameters of the cut legs. In general after fermionic integration a complicated sum over explicit amplitudes is generated. Here the focus is only on tracing the dependence on the parameter $t$. There are three distinct sources of t-dependence: the supersymmetry transformations, the fermionic argument of the last two amplitudes and the amplitudes themselves.

\subsubsection*{$t$ dependence from the supersymmetry transformation}
The $\xi$ supersymmetry transformation only depend on $t$ through their dependence on $\chi$ and $\overline{\chi}$. First examine $\chi^I$. There is a change of fermionic coordinates using two arbitrary independent spinors $\kappa$ and $\rho$ (which could be taken to be $1$ and $2$) for which
\begin{align}
\chi^{I,\dalpha} & = \frac{\lambda_1^{\dalpha} \iota_2^I - \lambda_2^{\dalpha} \iota_1^I}{\sbraket{\lambda_1 \lambda_2}} \\
                 & = \frac{\kappa^{\dalpha} \iota^I_2(t) - \rho^{\dalpha} \iota^I_1(t)}{\sbraket{\kappa \rho}}
\end{align}
By equating the two lines of the equation above one obtains explicit expressions for the new fermionic variables
\begin{align}
\iota^I_1(t) & = \frac{ \sbraket{\kappa \lambda_1} \iota^I_2 - \sbraket{\rho \lambda_2}  \iota^I_1}{\sbraket{\lambda_1 \lambda_2} }\\
\iota^I_2(t) & = \frac{ \sbraket{\rho \lambda_1} \iota^I_2 - \sbraket{\kappa \lambda_2}  \iota^I_1}{\sbraket{\lambda_1 \lambda_2} }
\end{align}
The same transformation can be performed for $\overline{\chi}$. In terms of the new variables the supersymmetry variations $\chi^I$ and $\overline{\chi}^I$ are $t$ independent. The transformation does however give a $t$ dependent Jacobian factor
\begin{equation}
(dF)_1 (dF)_2 \rightarrow \left(\frac{\sbraket{\kappa \rho}}{\sbraket{\lambda_1 \lambda_2}}\right)^4 (dF(t))_1 (dF(t))_2
\end{equation}
by the rules of fermionic integration: the integration is like a differentiation for which the usual chain rule combined with anti-symmetry of the coordinates gives the Jacobian. To get to the above equation the Schouten identity is employed. Since
\begin{equation}
\sbraket{\lambda_1 \lambda_2} = \frac{\sbraket{\lambda_1 1} \sbraket{2 \lambda_2} - \sbraket{\lambda_1 2}\sbraket{1 \lambda_2}}{\sbraket{12}}
\end{equation}
the Jacobian factor diverges as
\begin{equation}
J(t) \sim t^{4} + \mathcal{O}(t^3)
\end{equation}
exactly as in non gauge-symmetry broken field theory. This wraps up the discussion of $t$ dependence in the supersymmetry transformation.

The analysis of this subsection has also brought the $t$ dependence in the last two amplitudes in equation \eqref{eq:cohstatinttriangle} under control. The arguments there read
\begin{equation}
\pm m_3 \frac{\sbraket{\overline{\chi} 2}}{\sbraket{2 \lambda_3}} \qquad \textrm{and} \qquad \pm \qquad m_3 \frac{\sbraket{2 \chi}}{\sbraket{2 \lambda_3}}
\end{equation}
Through the fermionic change of variables the numerator of these is now $t$ independent, while the denominator is linear in $t$. For finite $m_3$ these terms therefore tend to zero in the limit.

\subsubsection*{$t$ dependence in the amplitudes}
The remaining $t$-dependence is now concentrated in shifts of the three amplitudes in \eqref{eq:cohstatinttriangle}. With the remark of the previous paragraph taken into account, these amplitudes now have definite quantum numbers on all legs. Two of these are BCFW type shifts for the massive vector bosons with spins $(+-)$, while the third is a $(--)$ shift of a slightly different type. By the relation of the broken Yang-Mills theory to its higher dimensional cousin discussed in section \ref{sec:fieldtheorypov} the same analysis as in \cite{ArkaniHamed:2008gz} applies here, because the large $t$ behavior of the spontaneously broken amplitude can be obtained from the higher dimensional Yang-Mills analysis in \cite{ArkaniHamed:2008yf}. In particular, the same result for the large momentum behavior of the two particle current applies,
\begin{equation}\label{eq:largetexp}
M^{ab}(t) = t g^{ab} h_0\left(\frac{1}{t}\right) +  A^{ab} h_1\left(\frac{1}{t}\right) + (a_4^a K^b + K^a a_4^b) h_2\left(\frac{1}{t}\right) + \frac{1}{t} B^{ab} h_3\left(\frac{1}{t}\right) + \mathcal{O}\left(\frac{1}{t}\right)
\end{equation}
with polynomial functions $h_i\left(\frac{1}{t}\right)$. The amplitude is obtained from this current by contracting with the polarization vectors and taking the on-shell limit.

From \eqref{eq:polvecsmas} in the large $t$ limit the polarization vectors are proportional to
\begin{equation}
e_i^{+} \sim \frac{1}{t} 2_{\alpha} 2_{\dalpha} \qquad  e_i^{-} \rightarrow t
\end{equation}
for \eqref{eq:largetspinorsI} and
\begin{equation}
e^{-} \sim \frac{1}{t} 1_{\alpha} 1_{\dalpha} \qquad  e_i^{+} \rightarrow t
\end{equation}
for \eqref{eq:largetspinorsI}. This is enough to prove through equation \eqref{eq:largetexp} that
\begin{equation}
A_{(--)} \rightarrow \frac{1}{t^3}
\end{equation}
while
\begin{equation}
A_{(+-)} \rightarrow \frac{1}{t}
\end{equation}
Putting everything together as
\begin{equation}
\lim_{t \rightarrow \infty} \left(t^4 \right) \left(\frac{1}{t^3} \right) \left(\frac{1}{t} \right) \left(\frac{1}{t} \right) = 0
\end{equation}
it is seen the triangle coefficients vanish. This completes the proof of the no-triangle property in $\mathcal{N}=4$ away from the origin of the moduli space through on-shell methods. Note in particular that the argument above applies to triangles with all massless, all massive or mixed legs in spontaneously broken $\mathcal{N}=4$. Hence the exact symmetry breaking pattern as parameterized by a single complex scalar is of no importance for the no-triangle property in the on-shell derivation.

\subsection{Off-shell}
The no-triangle property can also be proven by a modification of the original off-shell proof for $\mathcal{N}=4$ in \cite{Bern:1994zx} (see also \cite{Dixon:1996wi} and references therein). This used the background field method (see e.g. \cite{Abbott:1981ke}) to examine the loop integrals in a particular gauge. Within the background field method the gauge within the loop can be disentangled from the gauge in the trees. Hence loop effects can be calculated as \emph{gauge invariant} contributions to the effective action. In the following the map between spontaneously broken theories in four and unbroken theories in six dimensions will be used. Although this will not be done here, it should also be possible to set up the following argument in the broken four dimensional theory directly by modifying the techniques in \cite{Abbott:1981ke} for calculating scattering amplitudes to include constant background fields.

In six dimensions these contributions for the different particle species in the loop read
\begin{equation}\label{eq:contritoeffac}
\mathcal{L}_{\textrm{eff}} = \mathcal{L}_{\textrm{tree}} + 4 \mathcal{L}_{\textrm{scalar}} + 2 \mathcal{L}_{\textrm{chiral fermion}} + \mathcal{L}_{\textrm{gluon}}
\end{equation}
with
\begin{align}
\mathcal{L}_{\textrm{scalar}} & = \log \det\!\!\!\phantom{|}^{-1}_{s=0} \left(D_{\mu} D^{\mu} \right)_{D=6-2 \epsilon} \\
\mathcal{L}_{\textrm{chiral fermion}} & =  \frac{1}{2} \log \det\!\!\!\phantom{|}^{\frac{1}{2}}_{s=\frac{1}{2}} \left(D_{\mu} D^{\mu} + \sigma_{\mu\nu} F^{\mu\nu} \right)_{D=6-2 \epsilon} \\
\mathcal{L}_{\textrm{gluon}} & = \log \det\!\!\!\phantom{|}^{-\frac{1}{2}}_{s=1} \left(D_{\mu} D^{\mu} +  \Sigma_{\mu\nu} F^{\mu\nu} \right)_{D=6-2 \epsilon}  + \log \det\!\!\!\phantom{|}^1_{s=0} \left(D_{\mu} D^{\mu} \right)_{D=6-2 \epsilon}
\end{align}
Here $D_{\mu}$ is the covariant derivative for the external field $A$, $\sigma$ is the Lorentz generator in the spinor representation ($\sigma_{\mu\nu} = \frac{1}{4} [\gamma_\mu, \gamma_\nu]$) and $\Sigma$ is the Lorentz generator in the vector representation. The determinants are calculated with a particle of a certain spin in the loop. In the gluon loop the ghost contribution has been included. Written as a path integral the gluon contribution reads
\begin{equation}
\int [da_{\mu}] \exp \tr \left(\int d^{6-\epsilon}\, x \,\, a_{\mu}\left(g^{\mu\nu} D^2 +  (\Sigma_{\rho\sigma} F^{\rho\sigma})^{\mu\nu} \right)a_{\nu}       \right)
\end{equation}
Note the close similarity to the starting point of \cite{ArkaniHamed:2008yf} from which equation \eqref{eq:largetexp} was derived. This equation was crucial in the above on-shell construction.

The first thing to note now is that the above contribution to the effective action is not the same as the four dimensional theory we are interested in. The difference is that for the four dimensional theory the loop momentum integral in the off-dimensional directions needs to be frozen. However, the integrands are the same in the four and six dimensional cases and this is in fact the only thing needed here. More precisely, the higher dimensional background field method provides the natural background field gauge for the broken theory to do the power counting. The background version of the Lorentz gauge in higher dimensions for instance,
\begin{equation}
\frac{1}{\xi^2} \left(D_{\mu} a^{\mu} \right)^2
\end{equation}
for a quantum field $a$ with background field derivatives $D$ descends neatly to the lower dimensional theory as a definite, although intricate gauge choice. It remains to be shown that the integrand only carries a maximum of $n-4$ powers of the loop momentum in the numerator for the maximally supersymmetric Yang-Mills in four dimensions. That this happens in the higher dimensional Yang-Mills theory in the background field gauge is known, for completeness the full argument will be presented below.

The only dependence on the loop momentum enters in the covariant d'Alembertian, $D^2$. The terms with the maximal power of the loop momentum in the denominator follow from this term only, but it is not too hard to see that these terms will cancel within both the six dimensional susy matter multiplet as well as the six dimensional susy gauge multiplet. This follows because
\begin{equation}
4 \tr_{s=0}\left(1\right) - \frac{1}{2} \tr_{s=\frac{1}{2}}\left(1\right)  = 4 - \frac{1}{2} 8 = 0
\end{equation}
and
\begin{equation}
\frac{1}{2} \tr_{s=\frac{1}{2}}\left(1\right) - \tr_{s=1}\left(1\right) + 2 \tr_{s=0}\left(1\right)    = \frac{1}{2} 8 -6 + 2 = 0
\end{equation}
respectively. The terms with one power of the field strength are one down in power of the loop momentum and are proportional to
\begin{equation}
\tr\left(\sigma \right) = \tr \left(\Sigma \right) = 0
\end{equation}
Terms with two powers of the field strength arise from both the two fermions and the gluon loop. These terms are (in D dimensions) proportional to
\begin{equation}
\tr_V\left(\Sigma^{\mu\nu} \Sigma^{\rho\kappa} \right) = \frac{8}{D_{\textrm{spin}}} \tr_S \left(\sigma^{\mu\nu} \sigma^{\rho\kappa} \right) = 2 \left(g^{\mu\rho} g^{\nu \kappa} - g^{\mu\kappa} g^{\nu \rho} \right)
\end{equation}
where the subscript of the trace refers to the representation of the Lorentz group being traced and $D_{\textrm{spin}}$ is the dimensionality of the gamma matrix algebra. This leads to an exact cancelation again, when taking into account that there are two fermions. Note the neat trade-off of `numbers of fermions' with the spinor dimension in the above formula which shows the same cancelation works in $D=10$ for instance.

For the terms with three powers of the loop momentum the result that
\begin{align}
\tr_V \left(\Sigma^{\lambda \sigma} \Sigma^{\mu\nu} \Sigma^{\rho\kappa} \right) & = \frac{8}{D_{\textrm{spin}}} \tr_S \left( \sigma^{\lambda \sigma} \sigma^{\mu\nu} \sigma^{\rho \kappa} \right)
\end{align}
is needed. Again, the only important bit is the dependence on the spinor dimension which is traded with 'number of fermions' to produce the same cancelation as in four dimensions. This leads to the final sought-for cancelation.

Hence the integrand for the loop integrations in this particular setup display cancelations in the numerator down to $n-4$ powers of the loop momentum, compared to $n$ in the denominator. The standard reduction techniques to the integral basis maintain this discrepancy. This shows the entire amplitude can be expressed in terms of scalar (massive) boxes. In particular this immediately rules out tadpoles, bubbles and in particular triangles.

For the rational terms, note as in \cite{Badger:2008cm} there is a limited number of sources for rational terms to emerge in the expansion in terms of basis functions of equation \eqref{eq:massiveloopampexp}. In the four dimensional helicity scheme of treating the external polarizations the only sources are tensor box functions where two pairs of loop momenta are contracted with a metric and tensor triangle and bubbles where one pair of loop momenta is contracted. This implies there is a class of theories where the rational terms are absent: those for which the number of powers of the loop momentum in the numerator is always two down from the number of powers of the loop momentum in the denominator. The integral reduction procedure maintains this discrepancy. Hence in these theories only linear tensor triangles and bubbles and quadratic tensor boxes survive which can then be reduced further to the usual scalar functions. These do not yield enough powers of the loop momentum to yield rational terms. Hence there are no rationals in $\mathcal{N}=4$ anywhere on the moduli space.

\subsubsection*{Generalizations}
The `no-rationals' argument just described applies to a wide class of field theories: for instance to $\mathcal{N}=2$ coupled to matter on its Coulomb branch. This follows from the reasoning above as applied to the examples generated by the techniques in section \ref{sec:fieldtheorypov}. All these theories have two powers of loop momentum less in the numerator as compared to the denominator. In \cite{Bern:1994cg} field theories obeying the counting criterion where termed `cut-constructible'. It would be interesting to study supersymmetric theories where the gauge symmetry is broken by matter scalars (the Higgs branch) or with a lesser amount of supersymmetry. It is easy to conjecture that generically if an unbroken theory is cut-constructible the broken theory is as well.

\section{Discussion and conclusions}
Above it was shown that the vanishing of the triangle coefficients in maximal super Yang-Mills theory extends over the moduli space as parameterized by a single scalar vev through both an on as well as an off-shell method. There seem to be no fundamental obstructions to more general vacuum expectation values. The off-shell argument generalizes simply to $D=10$, while for the on-shell argument the coherent states constructed in \cite{Boels:2009bv} would be needed. In string theory this would correspond to multiple ($\geq 3$ ) parallel D-branes separated along different axis orthogonal to their world-volume. It is easy to see from the derivation that the on-shell and the off-shell method of proving no-triangles in $\mathcal{N}=4$ gauge theory are intimately connected through the background field method as a starting point. The main advantage of the off-shell method is that it applies directly to less supersymmetric theories as well and seems much less fragile than the on-shell method (which requires a certain number of choices along the way). The main advantage of the on-shell no-triangle argument in \cite{ArkaniHamed:2008gz} is that it applies to both maximally supersymmetric gauge theory as well as maximal supergravity.

Based on the results above an immediate theoretical question is whether a similar extension explored in this paper for the gauge theory is possible for $\mathcal{N}=8$ supergravity. At least the necessary BPS superspace is easily constructed. Furthermore maximal SUGRA is known to have a moduli space: this is controlled by the famed $E_{7,7}$ symmetry \cite{Cremmer:1978ds}. It seems physically reasonable to expect that turning on a vev for at least some of the scalars corresponds to moving unto the branch of BPS representations of the original $\mathcal{N}=8$ symmetry. The argument presented in this article could then be repeated for these BPS representations, leading again to `no triangles'. This would explicitly ignore the existence of another, new set of field coordinates in which a separate on-shell massless $\mathcal{N}=8$ representation is realized. The major difficulty of this calculation is the complication presented by the Lagrangian formulation of maximal SUGRA. Massive gravitons (let alone in $\mathcal{N}=8$) have after all a long and convoluted history, see for instance \cite{Chamseddine:2010ub} and references therein. It would be interesting to make this precise, as this would seem to show directly that the absence of triangles in the perturbation theory of maximal SUGRA is independent of the $E_{7,7}$ symmetry: the original supersymmetry would be enough. By extension this would indicate that possible finiteness of maximal SUGRA is not a consequence of this symmetry. This conclusion is also reached in \cite{Broedel:2009hu} by other means. As another avenue of attack on this problem perhaps the techniques of \cite{BjerrumBohr:2008ji} can be adapted to other points on the moduli space of $\mathcal{N}=8$ supergravity (and of $\mathcal{N}=4$ SYM). This would be interesting as their techniques have the advantage of applying to less supersymmetric gravity directly.

By the results of the off-shell derivation for the (in)possibility of rational terms, the amplitudes on the moduli space in $\mathcal{N}=4$ are one-loop scalar box only with the moduli space dependence confined to the box integrals and their coefficients. Therefore these amplitudes can easily be calculated by a quadruple cut, with the box coefficient neatly expressed by four fermionic integrals over four coherent state amplitudes. Moreover it is easy to speculate that at higher loops the exact same class of integral functions contribute as in the massless case, suitably extended to include massive propagators. This would all be a consequence of the conjectured exact dual conformal symmetry. As a first check of this the transformation of the coefficient of the box functions under dual and normal conformal transformations can be investigated. Note that by the connection between higher dimensional and Higgsed gauge theories uncovered in this article there is a direct connection between the Higgs regulator of \cite{Alday:2009zm} and standard dimensional regularization. It would be interesting to further explore the use of the BPS coherent states constructed explicitly in this article in tree and loop computations on the Coulomb branch of $\mathcal{N}=4$.

Although not the focus of the present article, the string theory picture of amplitudes under spontaneous symmetry breaking presented here seems very natural. It begs the question if more complicated D-brane setups relevant for for instance model building admit a similar reduction in terms of amplitudes.

\subsection*{Towards the standard model}
Apart from uncovering interesting structure in maximal super Yang-Mills part of the motivation for the present paper was to study the application of recent field theory ideas and techniques for massless four dimensional particles to the massive case directly. As a next step along this route it would be interesting to study examples of non-vanishing amplitudes. The ultimate goal here is of course the electroweak sector of the standard model. Although in the supersymmetric case the most straightforward symmetry breaking is via the adjoint in the longer run it will probably pay to start studying amplitudes first in the simplest supersymmetric setting. Note that by naive supersymmetric decomposition (i.e. state counting) it is seen that maximally supersymmetric one-loop amplitudes form at least a part of the non-supersymmetric amplitudes. Moreover, the `no rationals' results above for certain classes of susy theories in four dimensions indicate that these pieces can be calculated by much simpler means than appreciated before. This is however a story for another day.

\acknowledgments
The author would like to thank Simon Badger, Emil Bjerrum-Bohr, Johannes Henn, Niels Obers and Christian Schwinn in particular for comments and discussions. The figures were made with JaxoDraw \cite{Binosi:2008ig}. The research of RHB was supported by a Marie Curie European Reintegration Grant within the 7th European Community Framework Programme.

\appendix

\section{Shifting massive vector boson legs with generic spin axis}
\label{app:bcfwgenspinax}
In the main text the behavior of amplitudes with two massive vector boson legs under BCFW shifts was studied. This is summarized in table \ref{tab:BCFWkinmasvec} from the Feynman graph power counting and table \ref{tab:BCFWkinmasveccoulomb} from the higher dimensional interpretation of some Higgsed gauge theories. The analysis there is for a very special choice of quantum numbers for both legs. The objective in this appendix is to show that the same result as derived by Feynman graph power counting holds for a generic choice of polarization axis, common to both shifted particles. The main ingredient for this is the behavior of the polarization vectors for this choice under shifts.

As the scattering amplitudes are invariant under Lorentz transformations, it is enough to establish the result in one particular frame. In the frame in which
\begin{align}
k_i & = (E(m_i), 0, \ldots, k^i_3) \\
k_j & = (E(m_j), 0, \ldots, -k^j_3)
\end{align}
the vector $q$ (used to define the spinors which feed into the polarization vectors through equation \eqref{eq:decomponemom}) can be rotated by the residual symmetry in the center of mass frame to read
\begin{equation}
q = \left(q_0, q_1, 0, q_3 \right)
\end{equation}
with the constraint on the components that this vector is light-like. After the shift the form of the polarization vectors \eqref{eq:polvecsmas} for the massive vector bosons remains the same, but is now calculated with respect to the shifted momenta $k_i + z n$. Hence we need to obtain the shifted spinors from
\begin{equation}\label{eq:defofshiftedspinors}
\tilde{k}^i_{\alpha} \tilde{k}^i_{\dalpha} = k^i_{\alpha \dalpha} + z n_{\alpha \dalpha} - \frac{k_i^2}{2 q\cdot (k_i+ z n )} q_{\alpha} q_{\dalpha} \ .
\end{equation}
In the chosen frame the matrix on the right hand side of this equation has calculable eigenvectors and associated eigenvalues. One of these eigenvalues is zero by construction. To construct the $2$ two dimensional vectors on the left hand side of \eqref{eq:defofshiftedspinors} it is helpful to note that one of them is orthogonal to the zero eigenvalue eigenvector, while the other is proportional to the non-zero eigenvalue eigenvector. The non-zero eigenvalue fixes the proportionality factor. The expressions thus obtained are not very illuminating in general, although they do obey cross-checks with the mass-less limit. In the limit $z\rightarrow \infty$ these expressions simplify remarkably. In the center of mass frame in this limit the leading order behavior is given by
\begin{equation}
\begin{array}{ccccccc} \tilde{k}^i_{\alpha} & \rightarrow  & \left(\begin{array}{c} 0 \\ 1 \end{array} \right) & \quad \quad \quad & \tilde{k}^i_{\dalpha} & \rightarrow &  \left(\begin{array}{c} 2 z \\ k^i_0 - k_3 \end{array} \right) \\
\tilde{k}^j_{\alpha} & \rightarrow  & \left(\begin{array}{c} 2 z \\  k_0^j + k_3\end{array}  \right)  & \quad \quad \quad & \tilde{k}^j_{\dalpha} & \rightarrow &  \left(\begin{array}{c} 0 \\ 1\end{array} \right) \end{array}
\end{equation}
again up to the scaling ambiguity. Note that this scaling should not depend on $z$ as both $z=\infty$ and $z=0$ are not in $\mathbb{C}^*$ while the spinors should be well-defined on those points. In general the scaling ambiguity can make some limits of spinor expressions ill-defined if care is not taken with this point.

With the above behavior of the spinors for the shift under study, it follows that for massive vector bosons $i$ and $j$ having polarization $+$ and $-$ respectively with polarization vectors given in \eqref{eq:polvecsmas} these vectors behave
\begin{equation}
e_i^+ \sim e_i^- \sim \frac{1}{z}
\end{equation}
while for the opposite polarization choice for the same choice of shift,
\begin{equation}
e_i^- \sim e_i^+ \sim z \ .
\end{equation}
These observations can then be used to obtain the estimate of table \ref{tab:BCFWkinmasvec} through Feynman graph power counting and table \ref{tab:BCFWkinmasveccoulomb} from the higher dimensional interpretation of some Higgsed gauge theories for generic choice of common spin axis. It would be nice to have a covariant version of this argument.

\section{Deriving the on-shell superspace for $\mathcal{N}=4$ BPS multiplets} \label{app:susyrepsBPS}
In this appendix a coherent state representation of the BPS representations of the four dimensional supersymmetry algebra is presented. The coherent states can be interpreted as a function on an on-shell superspace. The approach is the four dimensional version of the general argument in \cite{Boels:2009bv} as applied to massless representations in six dimensions\footnote{See also \cite{Dennen:2009vk} for another view on the massless representations of the six dimensional supersymmetry algebra and \cite{4dmassivesusy} for the massive four dimensional case.}. The basic idea of the derivation is the same as in \cite{Grisaru:1977px}: the only thing needed to derive supersymmetric Ward identities is a covariant version of on-shell representation theory. These Ward identities can then be turned into on-shell superfields through coherent states as in \cite{ArkaniHamed:2008gz}. The non-trivial commutator of the supersymmetry algebra with central charges in four dimensions is usually written in terms of $D=4$ Dirac spinors as
\begin{equation}
\{Q_I, \overline{Q}_J\} = \gamma_{\mu} k^{\mu} \delta_{IJ} + \epsilon_{IJ} \left(Z_1 \id + \ii Z_2 \gamma_5\right)
\end{equation}
A linear transformation
\begin{equation}
Q = \frac{1}{\sqrt{2}}\left( Q_1 + \ii Q_2\right)
\end{equation}
and its conjugate bring the algebra to the form
\begin{equation}\label{eq:massbpsalg}
\{Q, \overline{Q}\}  = \gamma_{\mu} k^{\mu}+ \left(Z_1 \id + \ii Z_2 \gamma_5\right)
\end{equation}
and it's trivial companions. The right hand side can be recognized as the spin sum of solutions to the four dimensional Dirac equation albeit with a complex mass. This suggests to study solutions to the massive Dirac equation with a complex mass and definite spin. Note that equation \eqref{eq:massbpsalg} can also be recognized as the chiral projection of the supersymmetry algebra in six dimensions, providing the link to \cite{Boels:2009bv}. The necessary solutions to the four dimensional Dirac equation with definite spin have been obtained in \cite{Boels:2009bv} with a six dimensional motivation and read
\begin{equation}\label{eq:massspinsols}
\begin{array}{ccccccc}
u\left(k,\frac{1}{2}\right)  & =&  \left(\begin{array}{c} k^{\flat}_{\dalpha} \\ (Z_1 - \ii Z_2) \frac{q^{\alpha}}{\braket{q k^{\flat}}} \end{array} \right) & \quad & \overline{u\left(k,\frac{1}{2}\right)} &=& \left(\begin{array}{cc}   (Z_1 + \ii Z_2) \frac{q^{\dalpha}}{\sbraket{q k^{\flat}}} & k^{\flat}_{\alpha} \end{array} \right)\vspace{5pt} \\
u\left(k,-\frac{1}{2}\right) & = &\left(\begin{array}{c}   (Z_1 + \ii Z_2) \frac{q_{\dalpha}}{\sbraket{q k^{\flat}}} \\ k^{\flat,\alpha} \end{array} \right) & \quad & \overline{u\left(k,-\frac{1}{2}\right)} &=& \left(\begin{array}{cc}  k^{\flat,\dalpha} & (Z_1 - \ii Z_2)  \frac{q_{\alpha}}{\braket{q k^{\flat}}}  \end{array} \right)
\end{array}
\end{equation}
for the spinors and
\begin{equation}\label{eq:massconjspinsols}
\begin{array}{ccccccc}
u\left(q,-\frac{1}{2}\right) & = & \left(\begin{array}{c} q_{\dalpha} \\ 0 \\ 0 \end{array} \right) & \quad &  \overline{u\left(q,-\frac{1}{2}\right)} & = & \left(\begin{array}{ccc} 0 & 0& q_{\alpha}   \end{array} \right)\vspace{5pt} \\
u\left(q,\frac{1}{2}\right)  & = & \left(\begin{array}{c} 0 \\0 \\ q^\alpha  \end{array} \right) & \quad & \overline{u\left(q,\frac{1}{2}\right)} & = & \left(\begin{array}{ccc}  q^{\dalpha} & 0 & 0 \end{array} \right)
\end{array}
\end{equation}
for their conjugates. In these formulas a phase choice has been fixed by comparing to the massless limit. The spin is defined w.r.t. the spin axis given by the massless vector $q$. From these explicit solutions it is easy to see that
\begin{equation}
u\left(k,\frac{1}{2}\right), \qquad u\left(k,-\frac{1}{2}\right), \qquad  u\left(q,\frac{1}{2}\right), \qquad u\left(q,-\frac{1}{2}\right)
\end{equation}
forms a basis of Dirac spinors and
\begin{equation}
\overline{u\left(k,\frac{1}{2}\right)}, \qquad \overline{u\left(k,-\frac{1}{2}\right)}, \qquad  \overline{u\left(q,\frac{1}{2}\right)}, \qquad \overline{u\left(q,-\frac{1}{2}\right)}
\end{equation}
a basis for the conjugate spinors. Furthermore, the only non-vanishing spinor products are
\begin{equation}\label{eq:spinprodnonvan}
\begin{array}{cc}
\overline{u\left(k,\frac{1}{2}\right)} u\left(q,\frac{1}{2}\right)  = -  \overline{u\left(q,-\frac{1}{2}\right)} u\left(k,-\frac{1}{2}\right) & = \braket{k^{\flat} q} \\
\overline{u\left(k,-\frac{1}{2}\right)} u\left(q,-\frac{1}{2}\right)  = -  \overline{u\left(q,\frac{1}{2}\right)} u\left(k,\frac{1}{2}\right) & = \sbraket{k^{\flat} q}
\end{array}
\end{equation}
As anticipated above in this basis
\begin{equation}\label{eq:completenessspinbasis}
u\left(k,\frac{1}{2}\right) \overline{u\left(k,\frac{1}{2}\right)} + u\left(k,-\frac{1}{2}\right) \overline{u\left(k,-\frac{1}{2}\right)} = \gamma_{\mu} k^{\mu} +  \left(Z_1 \id + \ii Z_2 \gamma_5\right)
\end{equation}
holds with the $\gamma$ matrix in the chiral representation\footnote{A formula like \eqref{eq:completenessspinbasis} always exists, but the nice form here is a consequence of the phase convention in \eqref{eq:massspinsols} and \eqref{eq:massconjspinsols}.}, while of course also
\begin{equation}\label{eq:completenessspinbasisII}
u\left(q,\frac{1}{2}\right) \overline{u\left(q,\frac{1}{2}\right)} + u\left(q,-\frac{1}{2}\right) \overline{u\left(q,-\frac{1}{2}\right)} = \gamma_{\mu} q^{\mu}
\end{equation}
In view of equation \eqref{eq:completenessspinbasis} it is natural to expand the susy generators in the basis
\begin{equation}\label{eq:expansionofQ}
Q  \equiv u\left(k,\frac{1}{2}\right) Q_- + u\left(k,-\frac{1}{2}\right) Q_+ + u\left(q,\frac{1}{2}\right) \tilde{Q}_- + \left(q,-\frac{1}{2}\right) \tilde{Q}_+
\end{equation}
\begin{equation}\label{eq:expansionofbarQ}
\overline{Q} \equiv  \overline{Q}_+  \overline{u\left(k,\frac{1}{2}\right)}+  \overline{Q}_- \overline{u\left(k,-\frac{1}{2}\right)} +   \overline{\tilde{Q}}_+ \overline{u\left(q,\frac{1}{2}\right)} + \overline{\tilde{Q}}_- \overline{\left(q,-\frac{1}{2}\right)}
\end{equation}
Thanks to the known spinor products \eqref{eq:spinprodnonvan}, these expansions can easily be inverted,
\begin{equation}\label{eq:BPSgenerators}
\begin{array}{ccc}
Q_- & = & \frac{\overline{u(q,\half)} Q}{\overline{u(q,\half)}u(k,\half)}\\
Q_+ & = & \frac{\overline{u(q,-\half)} Q}{\overline{u(q,-\half)}u(k,-\half)}\\
\tilde{Q}_- & = & \frac{\overline{u(k,\half)} Q}{\overline{u(k,\half)}u(q,\half)}\\
\tilde{Q}_+ & = & \frac{\overline{u(k,-\half)} Q}{\overline{u(k,-\half)}u(q,-\half)}
\end{array} \qquad \qquad \begin{array}{ccc}
\overline{Q}_- & = & \frac{\overline{Q} u(q,-\half)}{\overline{u(k,-\half)}u(q,-\half)}\\
\overline{Q}_+ & = & \frac{\overline{Q} u(q,\half)}{\overline{u(k,\half)}u(q,\half)}\\
\overline{\tilde{Q}}_- & = & \frac{\overline{Q} u(k,-\half)}{\overline{u(q,-\half)}u(k,-\half)}\\
\overline{\tilde{Q}}_+ & = & \frac{\overline{Q} u(k,\half)}{\overline{u(q,\half)}u(k,\half)}
\end{array}
\end{equation}
Note that these generators are Lorentz invariant operators. Inserting equation \eqref{eq:completenessspinbasis} on the right hand side of \eqref{eq:massbpsalg} yields
\begin{equation}
\{Q, \overline{Q}\}  = u\left(k,\frac{1}{2}\right) \overline{u\left(k,\frac{1}{2}\right)} + u\left(k,-\frac{1}{2}\right) \overline{u\left(k,-\frac{1}{2}\right)}
\end{equation}
Using the spinor products \eqref{eq:spinprodnonvan} to project out on the left and right hand side of this equation yields the algebra of the generators of \eqref{eq:BPSgenerators}. The only non-trivial anti-commutators are
\begin{align}
\{Q_-, \overline{Q}_+\} &= 1 \\
\{Q_+, \overline{Q}_-\} &= 1
\end{align}
The remaining anti-commutators vanish and therefore in particular the generators which carry a tilde are represented trivially,
\begin{equation}
\tilde{Q}_-=\tilde{Q}_+=\overline{\tilde{Q}}_-=\overline{\tilde{Q}}_+ = 0
\end{equation}
Hence the representation theory of the algebra is equivalent to two copies of the fermionic harmonic oscillator. This is of course equivalent to the more usual `rest-frame' analysis, but with the added bonus that the generators above are Lorentz invariant: the analysis holds in any frame.  The analysis of the fermionic harmonic oscillator algebra is text-book material. The non-trivial generators raise and lower the spin quantum number with the spin axis determined by $q$ as indicated by the subscripts (this is straightforwardly verified in the rest-frame). Through \eqref{eq:expansionofQ} and \eqref{eq:expansionofbarQ} the action of a generic supersymmetry transformation on a state can be written down. This can be used to derive supersymmetric Ward identities as in \cite{Grisaru:1977px}. For the purposes of this article it will be useful to study these through the use of coherent states as in \cite{ArkaniHamed:2008gz}.

In the following the discussion will be specialized to $\mathcal{N}=4$ in four dimensions ($\mathcal{N}=(1,1)$ in $D=6$). This requires two copies of the susy algebra as analyzed above. For the purposes of this article two coherent states can be constructed using either the spin-up or spin-down mode of the massive vector boson. These read,
\begin{equation}
|\,\eta_I, \iota_I \rangle = e^{\sum_I \eta_I Q^I_{-} + \iota_I \overline{Q}^I_-}|\!\uparrow\,\rangle
\end{equation}
or
\begin{equation}
|\,\overline{\eta}_I, \overline{\iota}_I \rangle = e^{\sum_I \overline{\eta}_I Q^I_{+} + \overline{\iota}_I \overline{Q}^I_+}|\!\downarrow\, \rangle
\end{equation}
respectively. Here $I$ runs from one to two. The possibility of a coherent state representation follows simply from the fact that we are dealing with a representation of two copies of the fermionic harmonic oscillator. Definite states can be isolated by fermionic integration. The two representations are related by fermionic Fourier transform,
\begin{equation}
|\,\eta_I, \iota_I \rangle = \int (\overline{dF})  \,\, e^{\sum_I \left(\eta_I \overline{\eta}_I + \iota_I \overline{\iota}_I\right) } \,\, |\,\overline{\eta}_I, \overline{\iota}_I \rangle
\end{equation}
with the natural fermionic measure
\begin{equation}
(\overline{dF}) = d\overline{\eta}_{1} d\overline{\iota}_{1} d\overline{\eta}_{2} d\overline{\iota}_{2}
\end{equation}
More general coherent states can be obtained by Fourier transforming w.r.t. selected components. These states have other elements of the multiplet\footnote{These are the scalar or fermionic modes, as well as the longitudinal mode of the vector boson. Note that the $\mathcal{N}=4$ BPS multiplet can also be interpreted as a complex massive $\mathcal{N}=2$ multiplet.} as top-state, but will not be needed in this article. A general supersymmetry transformation is parameterized by Dirac spinor $\xi_I$ with an extra label (this would be six dimensional chirality) or its conjugate. Finite supersymmetry transformations act naturally on the coherent states as
\begin{align}\label{eq:trafosusycohstateDirac}
e^{\overline{\xi}_I Q^I} |\,\eta_I, \iota_I \rangle & = e^{\iota_I \left(\overline{\xi}^I u(k,-\half)\right)} \, \,|\,\eta_I + \left(\overline{\xi}_I u(k,\half)\right), \iota_I \rangle \\
e^{\overline{Q}_I \xi^I} |\,\eta_I, \iota_I \rangle & = e^{\eta_I \left(\overline{u(k,\half)} \xi_I \right)}\,\,|\,\eta_I, \iota_I + \left(\overline{u(k,-\half)} \xi_I \right) \rangle
\end{align}
and
\begin{align}\label{eq:trafosusycohstateIIDirac}
e^{\overline{\xi}_I Q^I} |\,\overline{\eta}_I, \overline{\iota}_I \rangle & = e^{\overline{\iota}_I \left(\overline{\xi}^I u(k,\half)\right) }\,\,|\,\overline{\eta}_I + \left(\overline{\xi}_I u(k,-\half)\right), \overline{\iota}_I \rangle \\
e^{\overline{Q}_I \xi^I} \,\,|\,\overline{\eta}_I, \overline{\iota}_I \rangle & = e^{\overline{\eta}_I \left(\overline{u(k,-\half)} \xi_I\right) }|\,\overline{\eta}_I, \overline{\iota}_I + \left(\overline{u(k,\half)} \xi_I\right) \rangle
\end{align}
These transformations (with spinor products written out in Weyl spinors) appear in the main text.

\section{Saving $\phi^4$ theory}\label{app:savingphi4}
Frequently in the literature and talks the statement is made that Yang-Mills theory is in a sense much better behaved than $\phi^4$ theory as that theory does not obey BCFW on-shell recursion relations. In \cite{Benincasa:2007xk}, appendix A it was shown how to `save' this theory by introducing auxiliary fields which allow BCFW-like recursion after performing \emph{two} shifts. This construction was also studied recently in \cite{Feng:2009ei}. It turns out one can do even better by treating the scalar field and its complex conjugate as two separate fields through the Lagrangian
\begin{equation}
\mathcal{L}  = \phi \Box \bar{\phi}  + \chi \Box \bar{\chi} + m_{\chi}^2 \chi \bar{\chi} - \sqrt{\lambda} m_\chi \chi \bar{\phi}^2 + \sqrt{\lambda} m_\chi \bar{\chi} \phi^2
\end{equation}
In the limit $\frac{m_{\chi}}{k_{\chi}} \rightarrow \infty$ the perturbation theory reproduces $\phi^4$ theory at the tree level. The point of this modification is that $\phi$ and $\bar{\phi}$ do not appear on the same vertex in the Feynman diagram expansion. Hence BCFW shifting a $\phi$ and a $\bar{\phi}$ field will always result in
\begin{equation}
A_n(z) \sim \mathcal{O}\left(\frac{1}{z} \right)
\end{equation}
behavior since there is always a propagator between the vertices with a shifted particle and the vertices themselves do not carry momentum. This proves BCFW recursion for the $\phi-\chi$ theory directly using this particular \emph{single} shift. A similar modification can be made to allow $\phi$ - $\phi$ shifts and their conjugate, thereby saving $\phi^4$ theory by a modification of the UV structure. Note that this observation is structurally the same as that made in \cite{Boels:2008fc}, appendix B for the Dirac-Born-Infeld action in four dimensions.

\addcontentsline{toc}{section}{References}

%%%%%%%%%%%%%%%%%%
\bibliographystyle{JHEP}

\bibliography{notriabib}
\end{document}